\documentclass[fleqn,usenatbib]{mnras}

\usepackage{times} 
\usepackage{listings}
\usepackage{graphics}
\usepackage{graphicx,xspace}
\usepackage{amsmath}
\usepackage{amssymb}
\usepackage{hyperref}
\usepackage{lscape}
\usepackage{rotating}
\usepackage{placeins}
\usepackage{natbib}
\usepackage{color}
\usepackage{txfonts}
\usepackage{mathptmx}
\usepackage{newtxtext,newtxmath}

\usepackage{bm}
\usepackage{subcaption}
\usepackage[dvipsnames]{xcolor}
\usepackage{caption}
\DeclareMathOperator{\erf}{erf}
\usepackage{grffile}

\newcommand{\ha}{H{\sc\,i}\xspace}
\newcommand{\dv}{{\rm d}}
\newcommand{\hw}{37}
\newcommand{\suffix}{pdf}
\newcommand{\vi}{\varv}

\title[Measuring the \ha mass function below the detection threshold]{Measuring the \ha mass function below the detection threshold}

\author[H. Pan et al.]{Hengxing Pan$^{1,2,3}$\thanks{E-mail: hengxing.pan@physics.ox.ac.uk},
Matt J.~Jarvis$^{3,4}$,
James R. Allison$^{3}$,
Ian Heywood$^{3,5,6}$,
\newauthor
Mario G. Santos$^{4,6}$,
Natasha Maddox$^{7}$,
Bradley S. Frank$^{6,8,9}$,
Xi Kang$^{1}$
\\
$^{1}$Purple Mountain Observatory, 2 West Beijing Road, Nanjing 210008, China\\
$^{2}$University of the Chinese Academy of Sciences, 19A, Yuquan Road, Beijing 100049, China\\
$^{3}$Astrophysics, University of Oxford, Denys Wilkinson Buiding, Keble Road, Oxford OX1 3RH, UK\\
$^{4}$Department of Physics and Astronomy, University of Western Cape, Cape Town 7535, South Africa\\
$^{5}$Department of Physics and Electronics, Rhodes University, PO Box 94, Grahamstown, 6140, South Africa\\
$^{6}$South African Radio Astronomy Observatory (SARAO), 2 Fir Street, Observatory, 7925, South Africa\\
$^{7}$Faculty of Physics, Ludwig-Maximilians-Universit\"at, Scheinerstr. 1, 81679 Munich, Germany\\
$^{8}$Department of Astronomy, University of Cape Town, Private Bag X3, 7701, Rondebosch, South Africa\\
$^{9}$Inter-University Institute for Data-Intensive Astronomy, Department of Astronomy, University of Cape Town, Private Bag X3, Rondebosch 7701, South Africa
}

\date{Accepted XXX. Received YYY; in original form ZZZ}

\pubyear{2019}

\begin{document}
\label{firstpage}
\pagerange{\pageref{firstpage}--\pageref{lastpage}}
\maketitle

\begin{abstract}
We present a Bayesian stacking technique to directly measure the \ha mass function (HIMF) and its evolution with redshift using galaxies formally below the nominal detection threshold.  We generate galaxy samples over several sky areas given an assumed HIMF described by a Schechter function and simulate the \ha emission lines with different levels of background noise to test the technique. We use {\sc Multinest} to constrain the parameters of the HIMF in a broad redshift bin, demonstrating that the HIMF can be accurately reconstructed, using the simulated spectral cube far below the \ha mass limit determined by the $5\sigma$ flux-density limit, i.e. down to $M_{\rm HI} = 10^{7.5}$ M$_{\odot}$ over the redshift range $0 < z < 0.55$ for this particular simulation, with a noise level similar to that expected for the MIGHTEE survey. We also find that the constraints on the parameters of the Schechter function, $\phi_{\star}$, $M_\star$ and $\alpha$ can be reliably fit, becoming tighter as the background noise decreases as expected, although the constraints on the redshift evolution are not significantly affected. All the parameters become better constrained as the survey area increases. In summary, we provide an optimal method for estimating the \ha mass at cosmological distances that allows us to constrain the \ha mass function below the detection threshold in forthcoming \ha surveys. This study is a first step towards the measurement of the HIMF at high ($z>0.1$) redshifts.
\end{abstract}


%%%%%%%%%%%%%%%%%%%%%%%%%%%%%%%%%%%%%%%%%%%%%%%%%%

%%%%%%%%%%%%%%%%% BODY OF PAPER %%%%%%%%%%%%%%%%%%

\section{Introduction}
Cold gas in the form of neutral hydrogen atoms (\ha) provides the reservoir from which molecules are formed and which subsequently go on to fuel star formation in galaxies, from the distant to the nearby Universe. To understand the whole picture of galaxy formation and evolution, it is crucial to know how the \ha gas evolves with cosmic time and as a function of environment. 

The preferred approach for tracing the \ha gas in the local Universe ($z \approx 0$) is via the direct detection of the neutral hydrogen 21 cm hyperfine emission line. Two prime examples of this are the HI Parkes All-Sky (HIPASS) Survey  \citep{barnes2001h, zwaan2005hipass}, conducted with the 64m Parkes radio telescope in Australia and the Arecibo Legacy Fast ALFA (ALFALFA) survey \citep{giovanelli2005arecibo, martin2010arecibo, jones2018alfalfa} conducted by the Arecibo radio telescope with  better sensitivity, angular and spectral resolution compared to the Parkes. However, both of these are limited in terms of redshift and sensitivity and have concentrated on surveying large swathes of the local Universe. Measuring the evolution of the \ha content of galaxies with redshift requires receivers that work to lower frequencies, coupled with high sensitivity and preferably with relatively high spatial and spectral resolution to avoid confusion \citep[e.g.][]{delhaize2013detection, jones2015stacking}.

However, we are entering an era of high sensitivity \ha surveys with the completion of a variety of new telescopes, in particular the wide-area surveys opened up with Phased Array Feed technology, principally the Australian Square Kilometre Array Pathfinder \citep[ASKAP; ][]{ASKAP-v2,DeBoer2009} and the Netherlands Aperture Tile In Focus \citep[AperTIF; ][]{APERTIF-HI,oosterloo2009apertif}, along with the extremely sensitive Meer Karoo Array Telescope \citep[MeerKAT; ][]{Jonas2009, Jonas2016}, which is conducting deeper but narrower blind surveys for \ha: The MeerKAT International GHz Tiered Extragalactic Exploration (MIGHTEE) Survey \citep{jarvis2017meerkat} and the Looking At the Distant Universe with the MeerKAT Array (LADUMA) survey \citep{Holwerda2011,LADUMA}. One of the most important goals of these surveys is to help us understand the cosmic evolution of \ha in the Universe \citep[e.g.][]{maddox2016optimizing,meyer2015advancing}.

However, even with the high sensitivity of MeerKAT, the 21 cm emission signal is so weak that only the most HI-massive galaxies will be directly detected beyond the local (z > 0.1) universe, unless very long integration times are used (e.g. \citealt{fernandez2016}). A proposed method to go further than the planned survey limits is to rely on stacking techniques, which use the known positions of galaxies from surveys at other wavelengths and allows the measurement of simple quantities, such as the average \ha mass (e.g. \citealt{rhee2018} and references therein).

Stacking in the context of {\ha} studies usually involves co-adding the spectral line data at the known locations of many galaxies to improve the signal-to-noise at the expense of information on individual galaxies contributing to the stack \citep[e.g.][]{delhaize2013detection,rhee2018}. Here, we instead present a Bayesian technique to model the distribution of {\ha} masses and not just co-add or average the data. This technique is based on a maximum-likelihood approach \citep{MitchellWynne2014}, which was extended into a fully Bayesian framework by \cite{zwart2015far}, who used it to model the source counts of faint radio continuum sources below the nominal detection threshold. Henceforth we simply describe this as Bayesian stacking, although we note that it is not strictly "stacking" in the normal sense of the word used in the literature.

Specifically, in this paper we extend this technique to determine the {\ha}-mass function (HIMF) directly from the {\ha}-line flux distribution constructed from extracting integrated fluxes across appropriate numbers of spectral channels at the redshift and position of galaxies selected in another waveband. We note that a similar technique has been used to measure the radio luminosity function of optically selected quasars below the noise \citep{Malefahlo2019}.

The idea behind this technique for {\ha} studies is that assuming we can determine the behaviour of the noise, the individual line-integrated 21-cm fluxes measured at the positions and redshifts of known galaxies can provide more information than just the average \ha mass. Accurate redshift information, in addition to positional information, is critical and so several spectroscopic surveys are planned to cover both the LADUMA and MIGHTEE fields \citep[e.g. the Deep Extragalactic VIsible Legacy Survey (DEVILS); ][]{DEVILS}.  Spectroscopic data from the Galaxy and Mass Assembly \cite[GAMA; ][]{GAMA} survey will provide acccurate redshift information for the ASKAP Deep Investigation of Neutral Gas Origins \citep[DINGO; ][]{DINGO} survey. 

We structure this paper as follows. In Section~\ref{sec:method}, we describe a source-count model for the \ha flux distribution and introduce a Bayesian stacking technique for constraining the HIMF model below the detection threshold. We also describe how we generate the \ha galaxy sample given an assumed HIMF model and how this is combined with realistic emission line profiles and input into a noisy spectral cube. In Section~\ref{sec:results}, we apply our method to the simulated data to test the robustness of our method in measuring the HIMF. In Section~\ref{sec:conclusions} we summarise our results, highlight some caveats in relation to simulated versus real data, and outline future work based on the real data from the MIGHTEE survey. We use the standard $\Lambda$CDM cosmology with a Hubble constant $H_{0}=67.7$ km$\cdot$s$^{-1}\cdot {\rm Mpc}^{-1}$, total matter density $\Omega_{\rm m}=0.308$ and dark energy density $\Omega_{\Lambda}=0.692$ \citep{ade2016planck}.

\section{Method}
\label{sec:method}
To simulate an \ha cube, we first need to construct the \ha flux distribution assuming a source-count model and the expected noise properties for a typical survey. 

\subsection{HIMF model}
\label{sec:himf}
Under the usual assumptions (e.g. optically thin gas), the \ha mass can be converted to the integrated flux  \citep[e.g.][]{meyer2017tracing} via
\begin{equation}
    M_{\rm HI} = 2.356 \times 10^5 D^2_L(1+z)^{-1} S,
	\label{eq:factor}
\end{equation}
where the \ha mass, $M_{\rm HI}$, is in solar masses, the luminosity distance to the galaxy, $D_L$, is in Mpc, and the integrated flux $S$ is in Jy\,km\,s$^{-1}$. The ($1 + z$) factor is needed when $S$ is expressed in units of Jy\,km\,s$^{-1}$ rather than Jy\,Hz.

For the HIMF we adopt a Schechter function model, which has been shown to fit the $z\sim 0$ HIMF from both HIPASS \citep{zwaan2005hipass} and ALFALFA \citep{jones2018alfalfa}, along with a pure density evolution term to characterise the evolution with redshift. Although a strong evolution of \ha mass is not expected from the little information we have, including this term does verify the flexibility of our approach. The specific form we adopt is therefore given by

\begin{equation}
    \phi(M_{\rm HI}, z) = \ln(10) \space \phi_\star\left(\frac{M_{\rm HI}}{M_\star}\right)^{\alpha + 1} e^{-\frac{M_{\rm HI}}{M_\star}} (1+z)^\beta,
    \label{eq:himf}
\end{equation}
where $\phi_\star$, $M_\star$, $\alpha$, and $\beta$ correspond to the normalization, characteristic mass, faint-end slope and power of the redshift evolution respectively. The HIMF represents the intrinsic number density of galaxies in the Universe as a function of their HI mass at a given redshift.
The adopted values of these parameters for our models are given in Table~\ref{tab:priors}.

We then define the volume-weighted average \ha mass function over the redshift bin defined by $z_1 < z< z_2$  by
\begin{equation}
    \Phi(M_{\rm HI}, z_1, z_2) =  \frac{ \int_{z_1}^{z_2} \frac{\dv V}{\dv z} \phi(M_{\rm HI}, z) \dv z }{  \int_{z_1}^{z_2} \frac{\dv V}{\dv z} \dv z},
    \label{eq:himf_z12}
\end{equation}
where $\dv V/\dv z$ is the differential co-moving volume at redshift $z$. The \ha mass density of the Universe $\Omega_{\rm HI}$ from $z_1$ to $z_2$ can then be estimated by integrating the $\Phi(M_{\rm HI}, z_1, z_2)$, which gives
\begin{equation}
    \Omega_{\rm HI} = \frac{1}{\rho_{\rm c}} \int_{\log_{10}(M_{\rm min})}^{\log_{10}(M_{\rm max})} M_{\rm HI} \Phi(M_{\rm HI}, z_1, z_2) \dv \log_{10}(M_{\rm HI}) ,
    \label{eq:omega_z12}
\end{equation}
where $\rho_{\rm c}$ is the critical density of the Universe at $z = 0$.

\begin{table}
	\centering
	\caption{The input parameters of the HIMF model used for our simulation of an \ha galaxy sample, along with the prior range used when running {\sc MultiNest}. $\phi_\star$ has units Mpc$^{-3}$ dex$^{-1}$, while the $M_\star, M_{\rm min}, M_{\rm max}$ are in units of M$_\odot$. For the input value of the cosmological HI density we integrate the input HIMF over the range $7.5 < \log_{10}(M_{\rm HI}/$M$_{\odot}) < 12.5$ and divide by the critical density (see equation~\eqref{eq:omega_z12}).}
	\label{tab:priors}
	\begin{tabular}{lrc}
	\hline
	Parameter & Input & Prior Probability Distribution \\
	\hline
	$\log_{10}(\phi_\star)$  & -2.318           & uniform $\in [-5, 0]$           \\
	$\log_{10}(M_\star)$     & 9.96             & uniform $\in [7, 12] $       \\
	$\alpha $                & -1.33            & uniform $\in [-5, 0] $           \\
    $\log_{10}(M_{\rm min})$ & 7.5              & uniform $\in [5, 10]$             \\
    $\log_{10}(M_{\rm max})$ & 12.5             & uniform $\in [8, 13] $      \\
    $\beta$                  & 1.0              & uniform $\in [-1, 4] $        \\
    \hline
	                               & 4.85& $0 < z < 0.1$  \\
	                               & 5.24& $0.1 < z < 0.2$  \\
	 $\Omega_{\rm HI} \times 10^4$ & 5.67& $0.2 < z < 0.3$  \\
	                               & 6.12& $0.3 < z < 0.4$    \\
	                               & 6.56& $0.4 < z < 0.5$    \\
	                               & 6.89& $0.5 < z < 0.55$   \\
	\hline
	\label{tab:priors}
	\end{tabular}
\end{table}

\subsection{Generating the galaxy sample}
Here we simulate what we expect from the MIGHTEE survey as an example of a future deep \ha survey. Using the HIMF described previously with the parameters shown in Table~\ref{tab:priors}, we generate a mass distribution of simulated galaxies over a volume similar to that expected for the MIGHTEE survey (i.e. $\approx 20$\,deg$^2$ over the redshift range $0<z<0.55$). The cube has a spatial resolution of $5$\,arcsec and a total of 19,504 26\,kHz channels from 913.4\,MHz to 1420.5\,MHz.

\begin{figure}
    \includegraphics[width=\columnwidth]{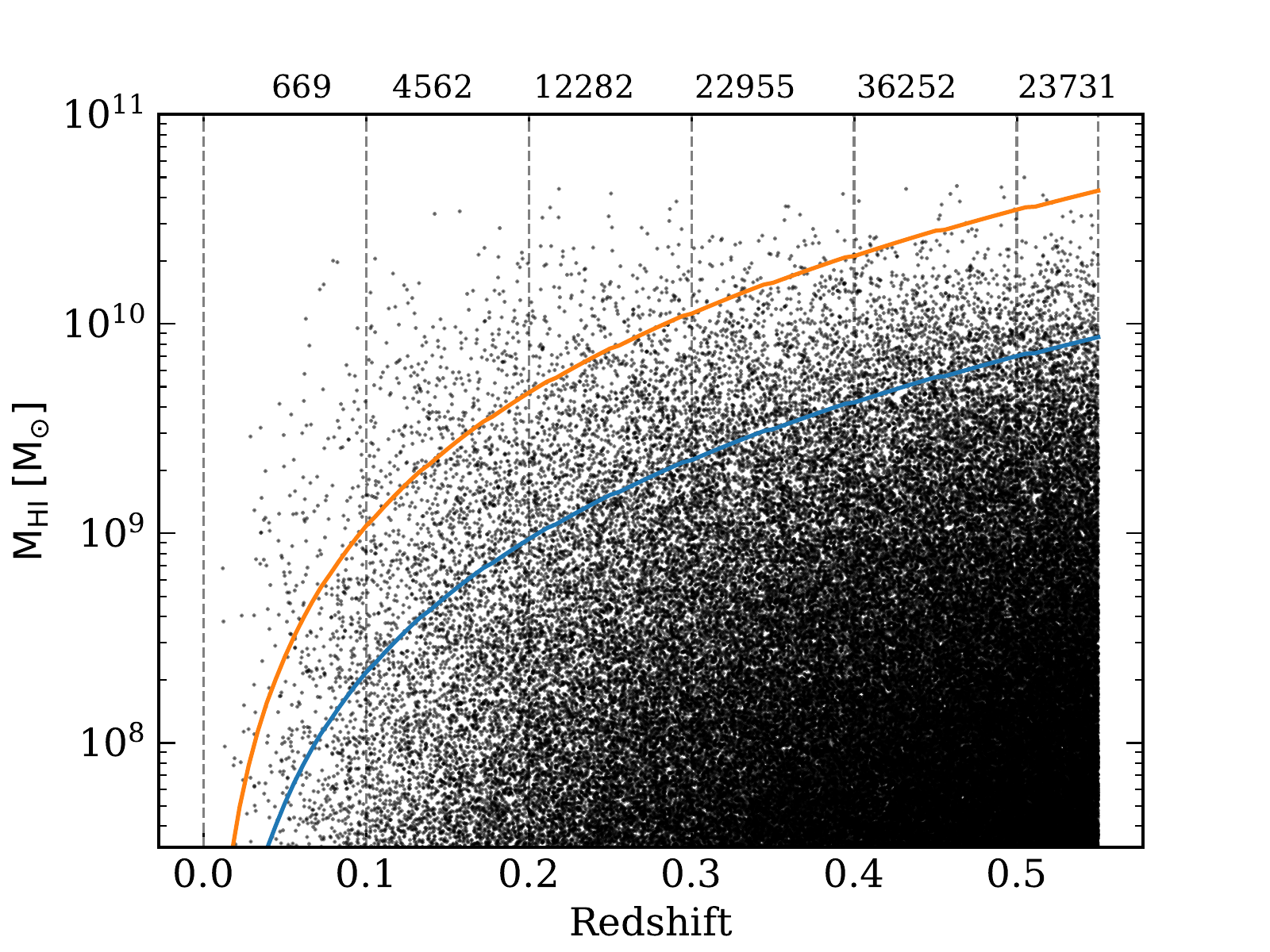}
    \caption{Simulated \ha mass versus redshift over a 1\,deg$^2$ area. The colour-coded lines indicate the $\sigma_{\rm n} = \sqrt{N_{\rm ch}}\sigma_{\rm ch}d\vi$ and 5$\sigma_{\rm n}$ detection threshold. The numbers listed at the top of the figure indicate the number of galaxies in each redshift bin.}
    \label{fig:mass_z}
\end{figure}

Fig.~\ref{fig:mass_z} shows the \ha mass against redshift for the generated galaxy sample. As we are aiming to explore the HIMF below the nominal noise threshold, the galaxy sample we use is volume-limited. However, we note that in reality we will require an input catalogue of galaxies in order to know where to extract fluxes from within the data cube, and the input catalogue itself will be flux limited. Therefore, any correlation that exists between the input catalogue selection and the \ha mass of the galaxies will effect our ability to measure the HIMF and we return to this in Section~\ref{sec:results}.

\subsection{Simulating \ha emission line profiles}

\begin{table}
\centering
\caption{The parameters of the busy function for simulating the \ha lines. Note that only parameters $b_{1}, b_{2}, c, x_{\rm p}$ and $w$ are randomly sampled, the centroid of the error function, $x_{\rm e}$, is fixed at the redshift of the source. $x_{\rm p}$ and $x_{\rm e}$ are in units of channel, while the line width has units km s$^{-1}$. }
\label{tab:busy}
\begin{tabular}{lll} 
\hline
Parameter &  Meaning & Probability Distribution \\
\hline
$b_{1}, b_{2}$ &steepness of line flanks & uniform $\in [0.3, 0.7] $              \\
$c\times 10^{5}$       &amplitude of the trough& uniform $\in [0.3, 0.7]$ \\
$|x_{\rm p}-x_{\rm e}|$    &difference of centroids & uniform $\in [0, 10] $\\
$2\times w$      &line width& uniform $ \in[55, 330]  $    \\
\hline
\end{tabular}
\end{table}

\begin{figure}
  \centering
    \includegraphics[width=0.8\columnwidth]{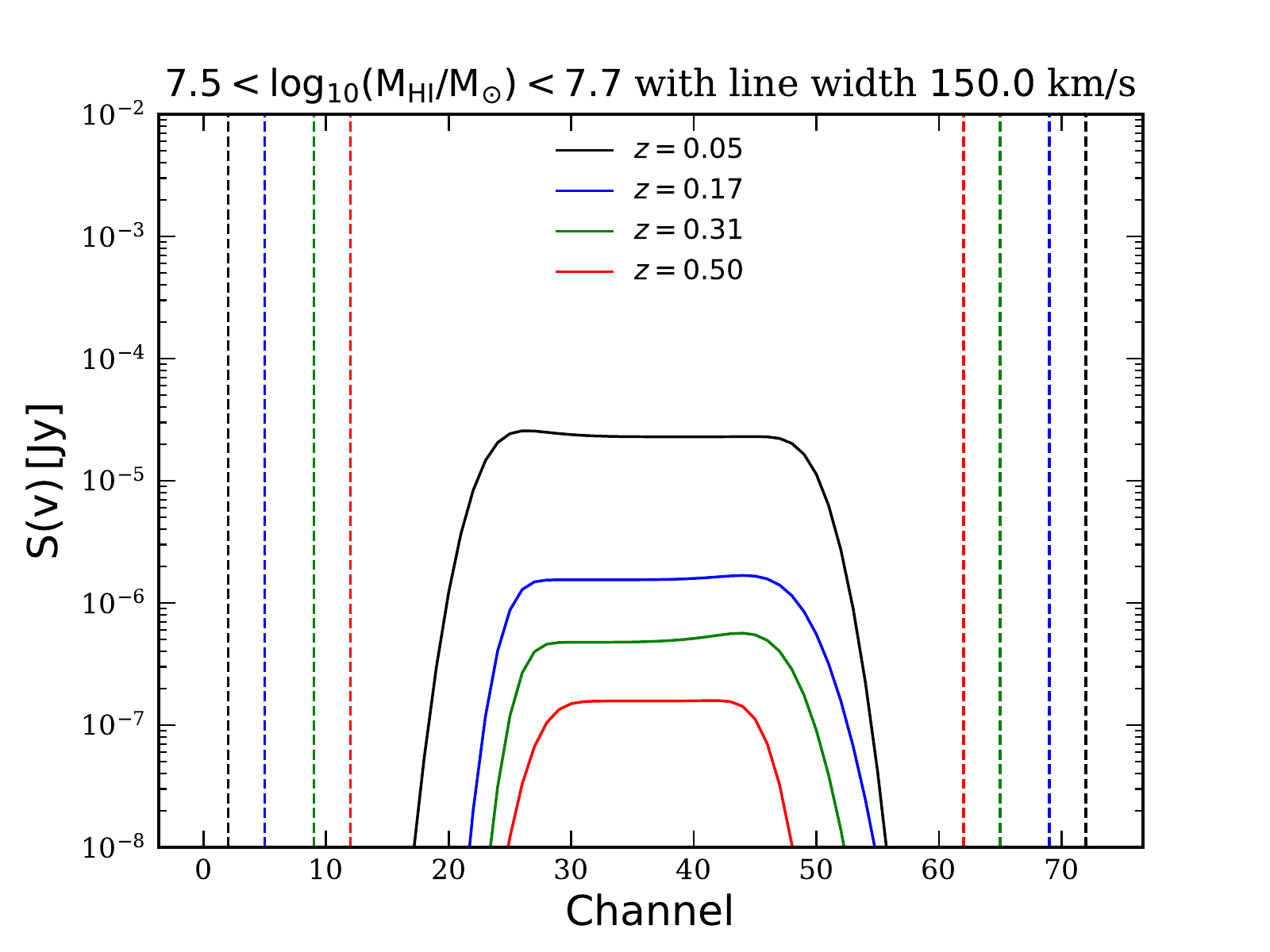}    
    \includegraphics[width=0.8\columnwidth]{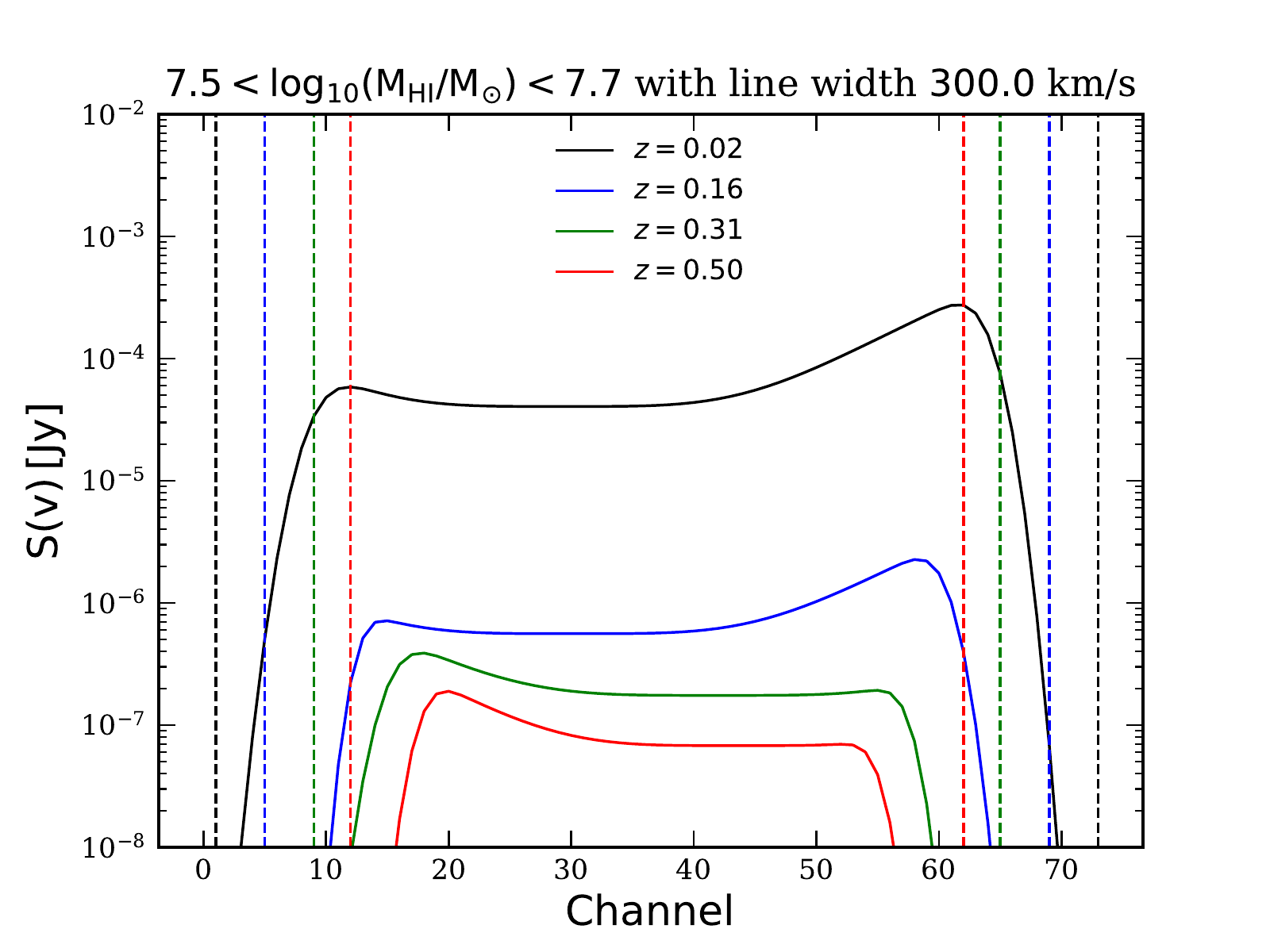}
    \includegraphics[width=0.8\columnwidth]{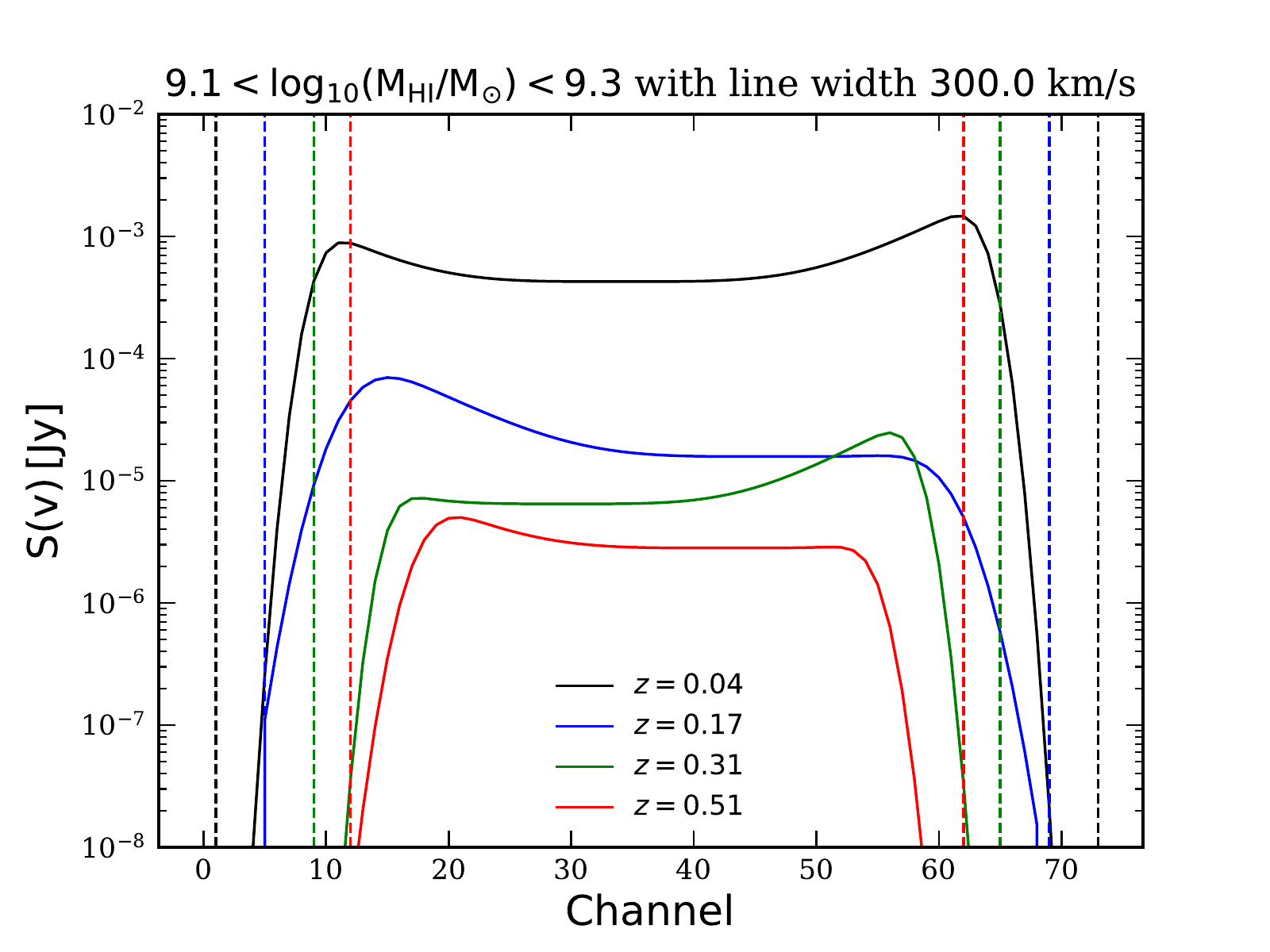}
 \caption{Flux density as a function of channel for galaxies with a mass of $\log_{10}(M_{\rm HI} / {\rm M}_\odot) = 7.5$ (top and middle panels) and $\log_{10}(M_{\rm HI} / {\rm M}_\odot) = 9.2$ (bottom panel). The colours indicate different redshifts. The vertical dashed lines show the limits of extracting the fluxes. Two different line widths are also shown $2\times w = 150$\,km\,s$^{-1}$ (top panel) and $300$\,km\,s$^{-1}$ (middle and bottom panels). The central channel of the spectra are fixed at channel 37 in these examples. }\label{fig:spectra}
\end{figure}

\begin{figure}
  \includegraphics[width=\columnwidth]{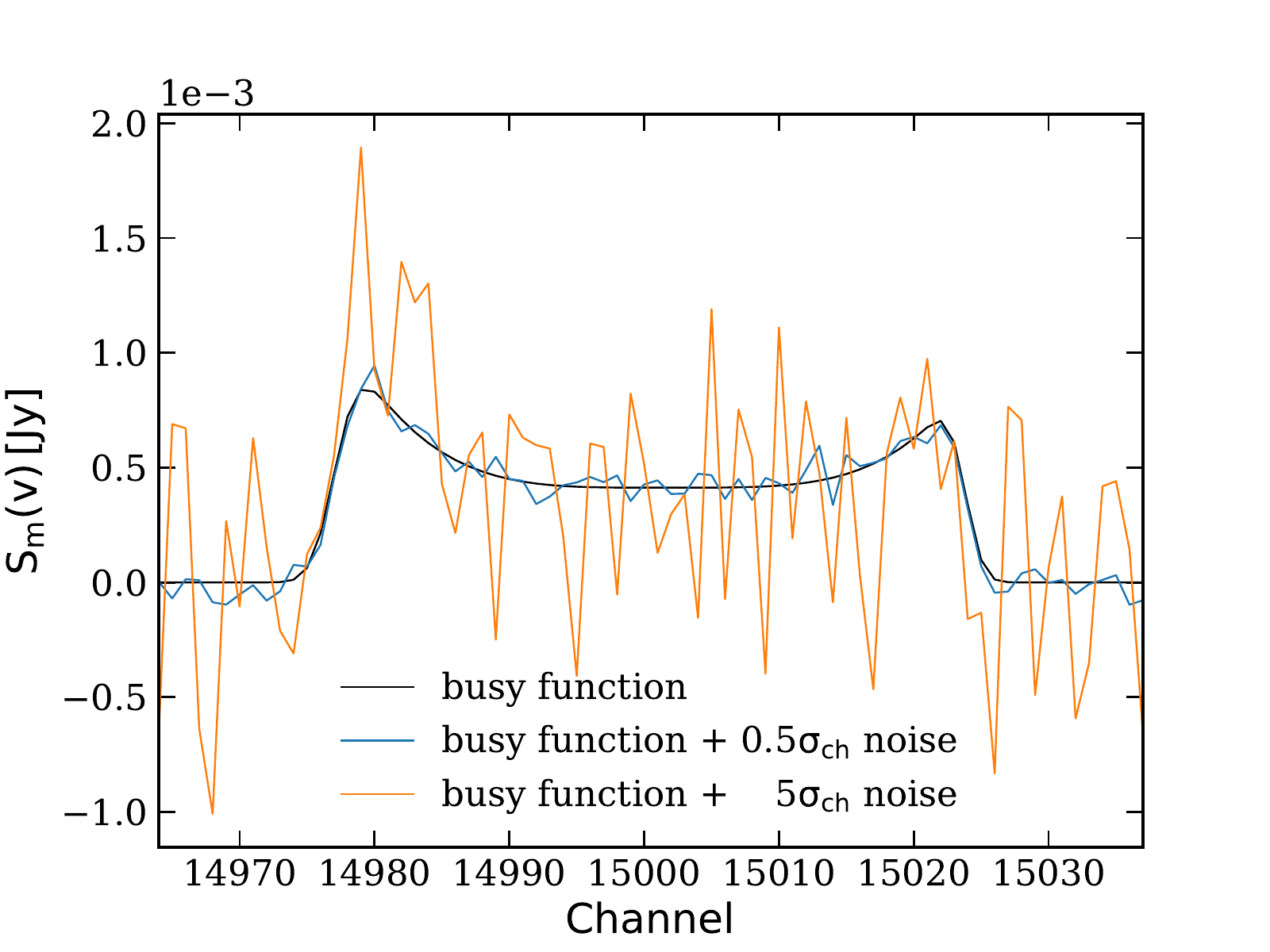}
  \caption{A simulated flux density as a function of channel generated by the busy function with random parameters. The black line shows the flux density without noise. The blue and orange lines indicate the flux density with $0.5\mathrm{\sigma_{ch}}, 5\mathrm{\sigma_{ch}} $ Gaussian noise.  }
  \label{fig:flux}
\end{figure}
Once the galaxy samples are generated, we simulate the flux density, $S_m(\vi)$, of \ha emission lines using the general form of the "busy function" with a fourth-degree polynomial trough \citep[i.e. equation (4) in ][]{westmeier2013busy}. The parameters $b_{1}$, $b_{2}$, $c$, $x_{\rm p}$ and $w$ are randomly sampled to assign a variety of line profiles to the \ha emission lines. In Fig.~\ref{fig:spectra} we show the flux density as a function of channel for galaxies of a given mass. The major variations are the steepness of line flanks, amplitude of the troughs, centroid of polynomials and widths of the profile. The difference between centroids of the error function and polynomial indicates the asymmetry of line profile, while the line widths approximate different galaxy inclinations. Although the form of lines changes, the integrated fluxes always keep constant for the sources with the same mass and redshift. The ranges used for these parameters are listed in Table~\ref{tab:busy}. We note that these ranges are set to be fairly broad rather than be realistic in order to prove the robustness of our approach. In particular,  a double-horn profile with the line width of 150\,km\,s$^{-1}$ for the low HI mass samples may not be expected from observations \citep[e.g.][]{ian2012AJ}, nevertheless one can find many Gaussian profiles easily from our simulations if looking at the flux density with narrower line widths.

We then generate the Gaussian noise with standard deviation similar to that expected from the MIGHTEE survey. Specifically we adopt $\sigma_{\rm ch} = 90$\,$\mu$\,Jy/channel, which is the expected noise per channel with a channel width of $26$\,kHz for the MIGHTEE surveys and $0.5\sigma_{\rm ch}$ as an indication of the improvement possible with a deeper survey, such as LADUMA. We also test the case for a much higher noise of $5\sigma_{\rm ch}$.
To determine the total flux density per channel across the spectral range, $S_m(\vi)$, we simply add the noise to the profile of the busy function and write the total flux density into the simulated cube. We note that only point sources are simulated in this work since our approach is aimed towards the high redshift regime where unresolved sources are expected to dominate given the spatial resolution of current telescopes.

In Fig.~\ref{fig:flux}, we show the simulated emission line as a function of the channel with $0.5\mathrm{\sigma_{ch}}$ and $5\sigma_{\rm ch}$ Gaussian noise. We do not know {\em a priori} the shape of the line profile,  we therefore fix the line limits that we measure the flux over at a velocity width  $N_{\rm ch}\times d\vi = 406$\,km\,s$^{-1}$ which corresponds to $N_{\rm ch} = 74$ at $z=0$ and $N_{\rm ch} = 48$ at $z= 0.55$. This ensures that we fully cover the range of expected line profiles. The integrated flux is then given by $S_m = \sum_{N_{\rm ch}} S_m(\vi) d\vi$, where $d\vi = 5.5$\,km\,s$^{-1}$ at $z=0$. We note that in our idealised case of uniform Gaussian noise $\sigma_{\rm ch}$,  the uncertainty associated with summing over all channels is given by $\sigma_{\rm n} = \sqrt{N_{\rm ch}}\sigma_{\rm ch}d\vi$ and is dependent on redshift (i.e. the $d\vi$ increases as $1+z$ and therefore the number of channels required to sample 406\,km\,s$^{-1}$ at the redshift of the source decreases as $1+z$ as shown in Fig.~\ref{fig:spectra}, thus the noise increases at $\sqrt{1+z}$). This is reasonable for our simulated cube as we already know that the noise follows a Gaussian distribution, but it might be problematic when we deal with real data, where the noise is likely to vary as a function of spectral window, and in such cases we would have to fit the noise distribution as a function of redshift. Furthermore, we would also not expect the signal-to-noise to be constant across the primary beam and return to this point in Section~\ref{sec:conclusions}.

With the known positions and redshifts from deep optical and/or near-infrared spectroscopy we are able to extract the flux across the number of channels that we expect to encompass the full \ha emission line and determine the source counts, which we subsequently model. Note that in this study we do not consider the effects of confusion since the new generation of deep \ha surveys will achieve $5-15$\,arcsec spatial resolution and have a spectral resolution of a few 10\,km\,s$^{-1}$. 
Indeed, if we consider how many "volumetric beams" (or voxels; where in this case a voxel is the beam area multiplied by the number of channels that we integrate over for the \ha line) per source we have given our choice of evolving mass function, then we estimate there are around 200 independent "voxels" per source, which is well above the nominal confusion level. We also note that \ha galaxies are not a strongly clustered population \citep[e.g.][]{Paperstergis2013}, and assume that lower-mass \ha galaxies are actually less clustered in general. As such our simulation of randomly positioned galaxies is probably not far from reality and would still be affected by confusion if it were an issue. This would result in us not accurately recovering the parameters of the input HIMF. However, we do note that close pairs of galaxies in groups will inevitably lead to some small level of confusion, but given the numbers involved this will be an insignificant perturbation on the derived parameters.

We also note that our method could be extended to incorporate confusion noise, either by building it into the model itself \citep[e.g.][]{Chen2017}, or via alternative methods of measuring the noise properties of the spectral cubes at the position of each source (Section~\ref{sec:conclusions}).

\subsection{The Likelihood Function}
\label{sec:Likelihood} 

Based on  Bayes' theorem, the probability of the model given the data is proportional to the likelihood function (i.e. probability of the data given the model). With the assumption that the number of sources in different flux bins are independent, the likelihood for all the flux bins is given by
\begin{equation}
    \mathcal{L} \propto \prod_{i=bin} P_i(k_i|\dv N/\dv S), 
	\label{eq:likeli}
\end{equation}
where $P_i(k_i|dN/dS)$ is the probability of obtaining $k_i$ objects in bin $i$ given a source-count model for the \ha flux distribution $\dv N/\dv S$ \citep[see e.g.][]{MitchellWynne2014,zwart2015far}.

The measured flux from a \ha emission line, $S_m$, is a combination of the intrinsic flux from the source, $S$ plus the contribution from the noise, $S_n$, which can obviously be positive or negative.

With the assumption that the number of sources in a given patch of sky follows a Poisson distribution, the probability of finding $k_i$ objects over the observed sky area in the $i^{th}$ flux bin [$S_{m_i},  S_{m_i} + \Delta S_{m_i}$] is given by 
\begin{equation}
    P_i(k_i|\dv N/\dv S) = \frac{\lambda_i^{k_i} e^{-\lambda_i}}{k_i!},
	\label{eq:prob}
\end{equation}
where $\dv N/\dv S$ is our source-count model for the \ha galaxies in the observed area and $\lambda_i$ is the theoretically-expected number of \ha sources in that bin.

If the noise follows a given distribution $P_n(S_n)$,  we can write  $\lambda_i$ as the convolution of $dN/dS$ and $P_n(S_n)$
\begin{equation}
    \begin{aligned}
    \lambda_i &= \int_{S_{min}}^{S_{max}} \dv S\frac{\dv N}{\dv S}\int_{S_{m_i}}^{S_{m_i}+\Delta S_{m_i}} \dv S_m P_n(S_m-S).
    \label{eq:lambda}
    \end{aligned}
\end{equation}
where $S_{\rm min}$ and $S_{\rm max}$ are the minimum and maximum flux that can be measured from the data. These correspond to $M_{\rm min}$ and $M_{\rm max}$ in our model via equation~\eqref{eq:factor}.

For the idealised case of where the noise follows a Gaussian distribution centered at zero with a variance $\sigma_{\rm n}^2$, equation~\eqref{eq:lambda} becomes
\begin{equation}
    \begin{aligned}
    \lambda_i &= \int_{S_{min}}^{S_{max}} \dv S\frac{\dv N}{\dv S}\int_{S_{m_i}}^{S_{m_i}+\Delta S_{m_i}} \dv S_{m} \frac{1}{\sigma_{\rm n}\sqrt[]{2\pi}} e^{-\frac{(S_{m}-S)^2}{2\sigma_{\rm n}^2}} \\
    		  &= \int_{S_{min}}^{S_{max}} \dv S\frac{\dv N}{\dv S}
              \frac{1}{2} \left[\erf\left(\frac{S_{m_i}-S}{\sqrt{2}\sigma_{\rm n}}\right)- \erf\left(\frac{(S_{m_i} + \Delta S_{m_i}-S}{\sqrt{2}\sigma_{\rm n}}\right)\right],
	\label{eq:salient}
    \end{aligned}
\end{equation} 
where $\sigma_{\rm n} = \sqrt{N_{\rm ch}}\sigma_{\rm ch}d\vi$, and $N_{\rm ch}$ is the number of channels over the flux density profile of the \ha line for each galaxy, $\sigma_{\rm ch}$ is the RMS per channel and $d\vi$ is the velocity width of each channel.

We then relate the source-count model with the redshift dependent HIMF, $\phi(M_{\rm HI}, z)$, by
\begin{equation}
    \begin{aligned}
        \frac{\dv N}{\dv S} &= \frac{\dv N}{\dv M_{\rm HI}} \frac{\dv M_{\rm HI}}{\dv S} \\
        &= \frac{\dv N}{\dv \log_{10}(M_{\rm HI})} \frac{\dv \log_{10}(M_{\rm HI})}{\dv M_{\rm HI}} \frac{\dv M_{\rm HI}}{\dv S} \\
        &=\int_{z_1}^{z_2} \frac{\dv V}{\dv z} \phi(M_{\rm HI}, z) \space  \frac{\log_{10}(e)}{M_{\rm HI}} 2.356 \times 10^5 D^2_L(1+z)^{-1}\dv z,
        \label{eq:dnds}
    \end{aligned}
\end{equation}
where $z_1$ and $z_2$ indicate the redshift limits for a given redshift bin and $\dv V = r^2 \Omega\dv r $ where $r$ is the co-moving distance and $\Omega$ is the solid angle of a given survey.

\subsection{Parameter estimation}

Here we describe our approach to constrain the parameters of the input mass function so that we compare the best-fitting parameters with the mass function parameters in order to test if we could recover them. 

\subsubsection{Priors}
\label{sec:priors}
Priors are the prior knowledge or limits of a parameter before any relevant evidence is taken into account.  In Bayesian statistical inference, they provide the sampling parameter space for computing the posterior probability. We assign a uniform prior probability distribution to the power terms $\alpha$ and $\beta$ and adopt uniform logarithmic priors for $\phi_{\star}$, $M_\star$, $M_{\rm min}$ and $M_{\rm max}$. These are listed in Table~\ref{tab:priors}.

\subsubsection{Multinest}
\label{sec:Multinest}
In order to determine the posterior probability distribution and fit the model, we use {\sc Multinest}\footnote{\url{https://github.com/JohannesBuchner/PyMultiNest}} to sample the prior parameter space \citep{feroz2009multinest, buchner2014x}. Multinest is an efficient and robust Bayesian inference tool based on a nested sampling technique \citep{skilling2004nested}, which produces the posterior samples from distributions with an associated error estimate and the Bayesian evidence term allowing model selection. In this paper we accept the median of posterior samples as the best estimate of each parameter considering the complex nature of the posterior, and the set of parameters with the maximum likelihood as the best fitting HIMF. We do not compare different models, we therefore do not use the evidence term. However, we note that an obvious use of the evidence would be to determine the model which best describes the form of the evolution of the HIMF, and we leave this to future work using real data.

\subsection{Survey Parameters}
\label{sec:surveys}
For our baseline survey we assume a 1\,deg$^{2}$ survey with noise of $\sigma_{\rm ch} = 90$\,$\mu$Jy/channel over the redshift range $0<z<0.55$. We then consider the effects of reducing and increasing the rms of the spectral line cube to investigate how the noise properties affect our ability to measure the HIMF accurately. Furthermore, increasing the background noise with a fixed extraction window is also similar to the situation where only photometric redshifts are available, as one has to extract a larger window for fully covering the line profile and this will inevitably integrate more noise into the measured flux. In real observations we are likely to be limited by an optical flux limit for the spectroscopic sample that is required in order to know where to extract fluxes in the spectral line cube. In reality, the supporting spectroscopy will be limited by a complex optical flux limit that is dependent on galaxy colour, morphology and emission line properties. For our purposes, we assume that this flux limit roughly corresponds to a stellar mass limit (although we note that we do not expect this to be a clean one-to-one relation) and many studies have shown that stellar mass and \ha mass are correlated, particularly at $M_{\rm HI} < 10^9$\,M$_\odot$ \citep[e.g. Fig.~1 in ][]{Maddox2015}. We therefore also investigate how a higher minimum \ha mass for our simulated galaxy sample may affect the recovered HIMF. Finally, we also investigate how the survey area influences the constraints by considering a range of surveys from $0.3 - 20$\,deg$^{2}$.

\section{Results}
\label{sec:results}

We start by looking at the reconstructed parameters of our baseline survey and then investigate the effects of increasing noise, minimum mass and survey area. Note that we include all simulated samples (i.e. $M_{\rm HI} > 10^{7.5}$ M$_{\odot}$) across the redshift range $0<z<0.55$ in the following analysis for exploring the effects of increasing noise and survey area, except for the effect of a higher minimum mass cut.

For analysing the redshift dependence of our approach, we simply split the samples into six individual redshift bins (z-bins) as shown on the top of Fig.~\ref{fig:mass_z}. Although there should be an optimal way of setting the bins (for instance, the same way as we do for binning the flux below) our current results will only be improved by a better redshift binning scheme.

To capture the shape of the flux distribution in an individual redshift bin accurately, it is sensible to set the flux bins to be uniform in log-flux space since the low flux sources dominate. However, considering that nearly half of low flux sources have negative measured fluxes (due to the noise), we instead apply the Bayesian Blocks method \citep{scargle2013studies} to determine suitable flux bins. The Bayesian Blocks method requires the minimization of a cost function across the data set and allows varying bin widths while being robust to statistical fluctuations, which is ideal for dynamic distributions of the fluxes. The user can determine how many bins are generated through the correct detection rate $p_0$ defined by equation (11) of \cite{scargle2013studies}. A small $p_0$ will be more robust to statistical fluctuations in the data, but could be overly coarse. The optimal value of $p_0$ is typically determined empirically. At $z > 0.1$ we accept the default value of $p_0 = 0.05$, but at $z < 0.1$ we use $p_0 = 0.5$ to avoid coarseness of the binning. In general, our results are insensitive to a large range of reasonable values for $p_0$.

\begin{figure*}
  \centering
  \begin{subfigure}[b]{0.5\textwidth}
    \includegraphics[width=\columnwidth]{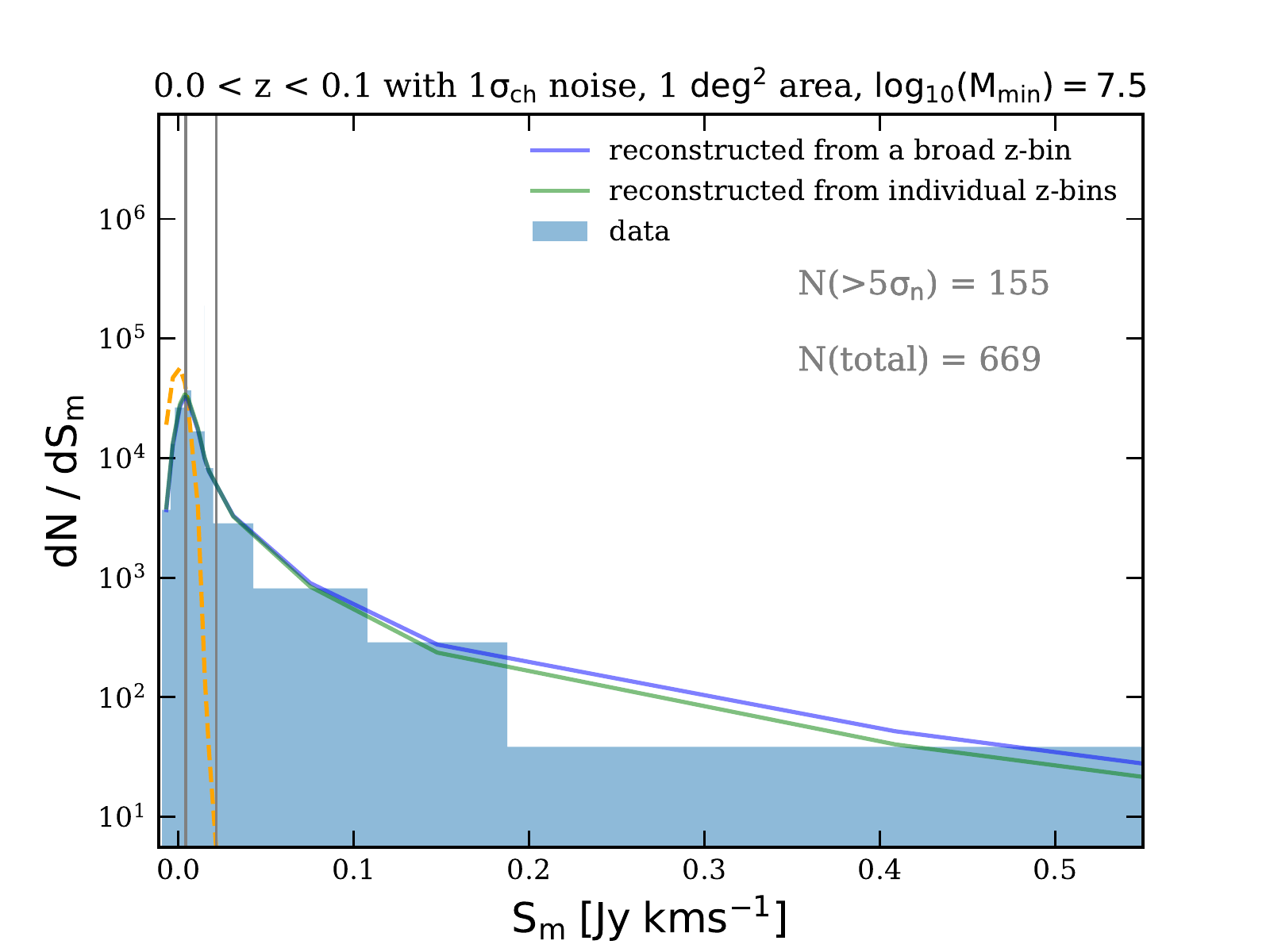}
    \includegraphics[width=\columnwidth]{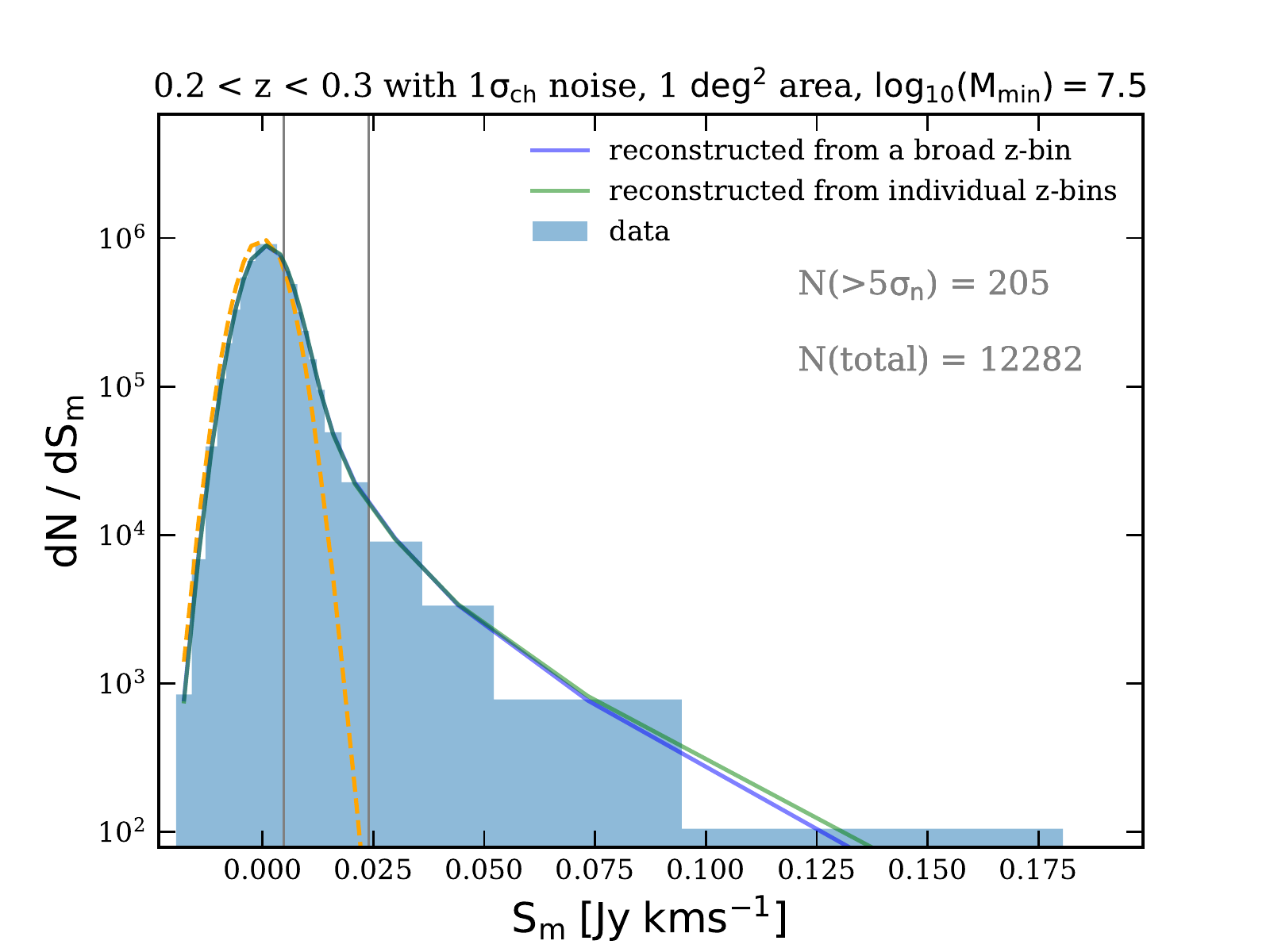}
    \includegraphics[width=\columnwidth]{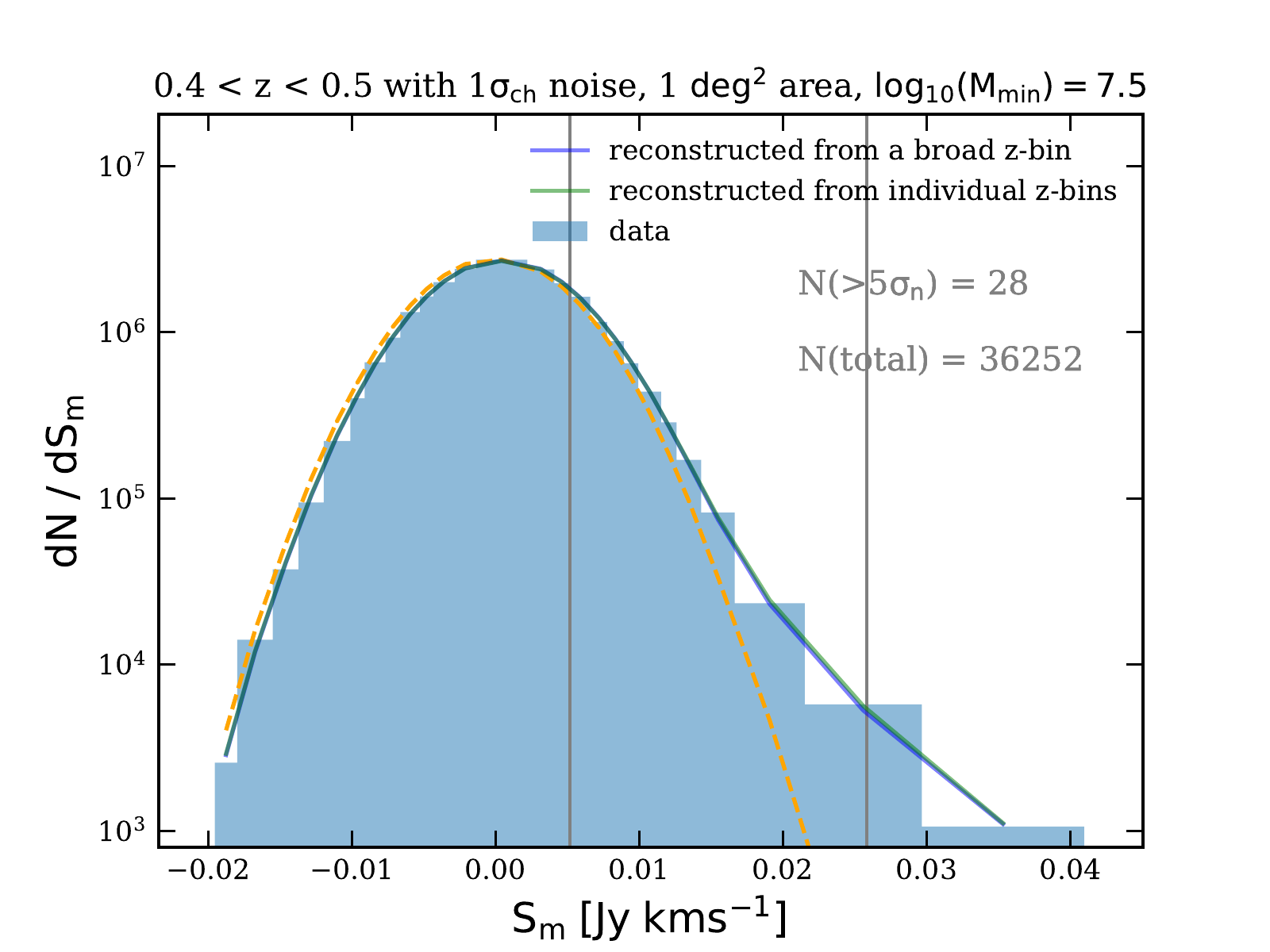}
  \end{subfigure}%
  \hfill
  \begin{subfigure}[b]{0.5\textwidth}
    \includegraphics[width=\columnwidth]{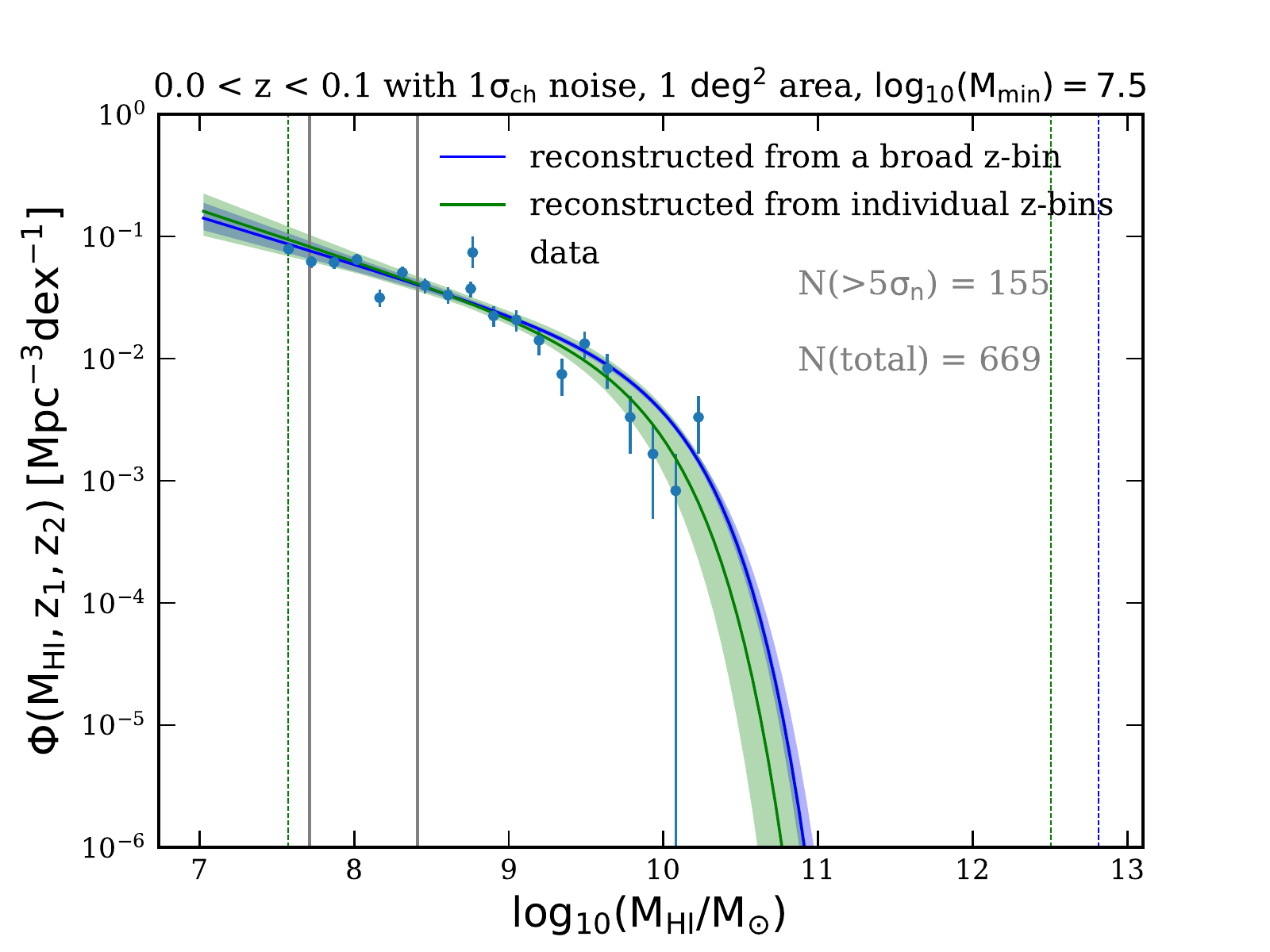}
    \includegraphics[width=\columnwidth]{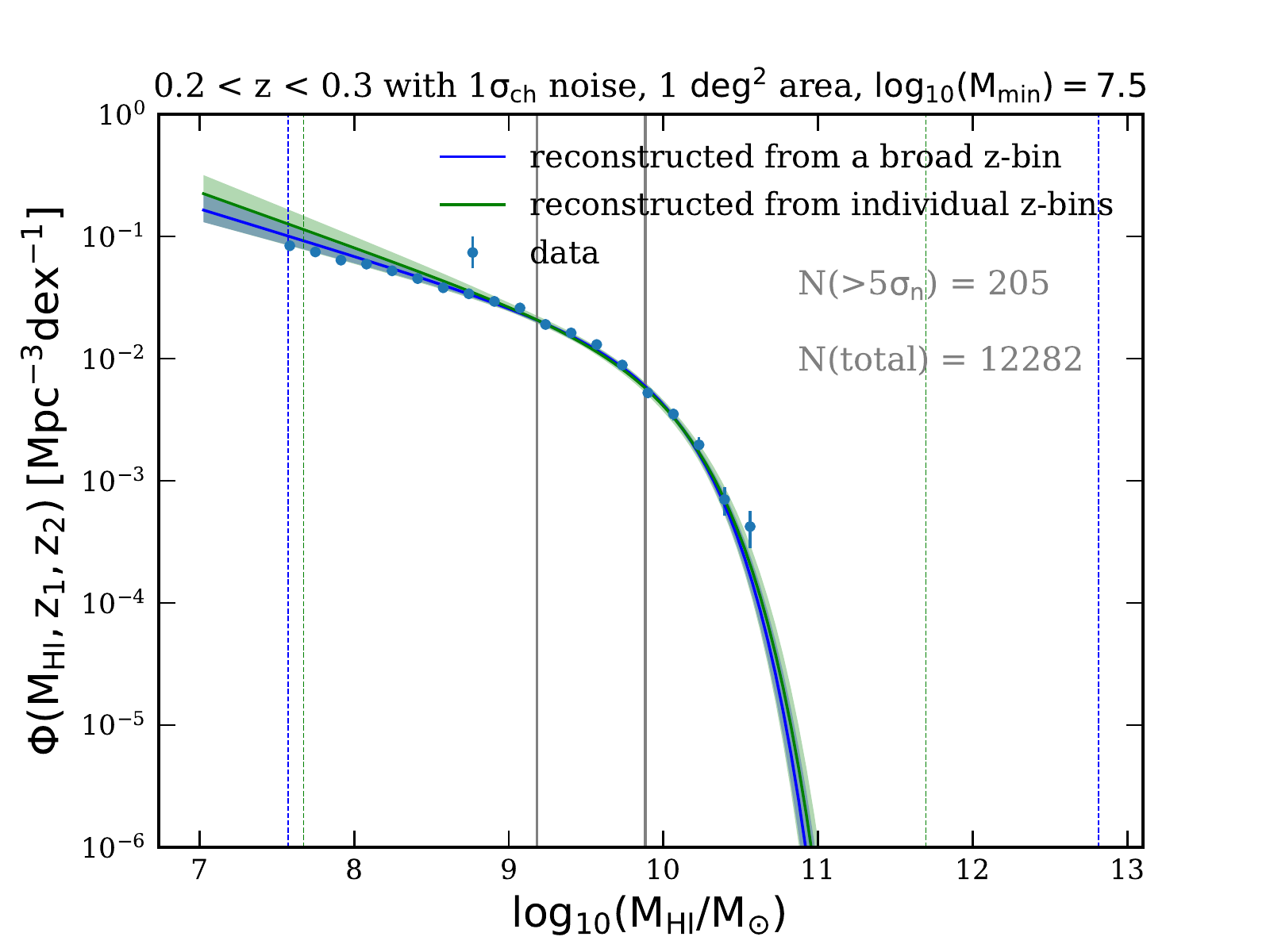}
    \includegraphics[width=\columnwidth]{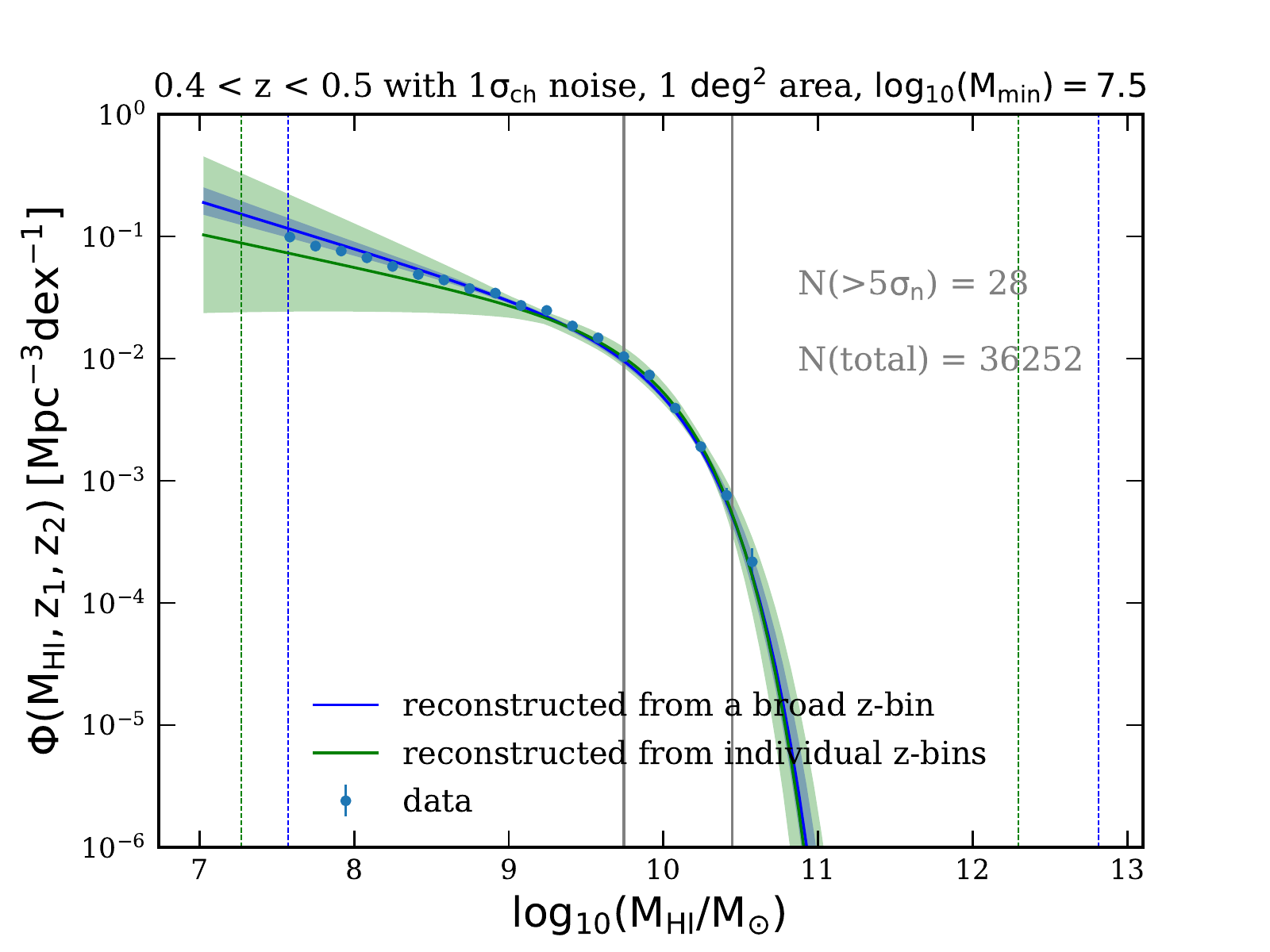}
  \end{subfigure}%
    \caption{Reconstructed flux $S_m$ distribution (left) and  HIMF (right) in different redshift bins from our baseline survey. 
    The blue lines show the results from a broad redshift bin ( i.e.  $0 < z < 0.55$) while the green lines show the results from an individual redshift bin. 
    The grey vertical lines indicate the $\sigma_{\rm n} = \mathrm{\sqrt{N_{\rm ch}}} \mathrm{\sigma_{ch}} d\vi$ and formal 5$\sigma_{\rm n}$ detection threshold at the centre of each redshift bin. Only three redshift bins are shown for simplicity. For the left panels, we apply the Bayesian Blocks method \citep{scargle2013studies} that allows varying bin widths to the data. The dashed orange line denotes a Gaussian distribution with a width equal to $\sigma_{\rm n}$. For the right panels, the color-coded regions are the 68\% credible intervals in the HIMF estimated from the posterior samples. The color-coded vertical lines are $\log_{10}(M_{\rm min})$ and $\log_{10}(M_{\rm max})$ of the best fitting from a broad redshift bin and an individual redshift bin. The errors on the simulated data points are just Possion errors. }
    \label{fig:hist_himf}
\end{figure*}

\renewcommand{\arraystretch}{1.4}

\begin{table*}
	\centering
	\caption{Reconstructed parameters of the HIMF model from our baseline survey with changing background noise, where we fix $\log_{10}(\phi_\star) = -2.318 $ for individual redshift bins.}
	\label{tab:pos_s}
	\begin{tabular}{lrrrrrrrr} 
	\hline
	\hline
	Parameter & $\times \; \sigma_{\rm ch} $ & $0 < z < 0.1$ & $0.1 < z < 0.2$ & $0.2 < z < 0.3$ & $0.3 < z < 0.4$ & $0.4 < z < 0.5$ & $0.5 < z < 0.55$  & $0 < z < 0.55$\\
	\hline
        $\log_{10}(\phi_\star)$ & 0.5 & -2.318 & -2.318 & -2.318 & -2.318 & -2.318 & -2.318 & -2.338$^{+0.026}_{-0.027}$ \\
$\log_{10}(\phi_\star)$ & 1.0 & -2.318 & -2.318 & -2.318 & -2.318 & -2.318 & -2.318 & -2.397$^{+0.039}_{-0.042}$ \\
$\log_{10}(\phi_\star)$ & 5.0 & -2.318 & -2.318 & -2.318 & -2.318 & -2.318 & -2.318 & -2.378$^{+0.131}_{-0.152}$ \\

	\hline
        $\log_{10}(M_\star)$ & 0.5 & 9.819$^{+0.077}_{-0.071}$ & 9.977$^{+0.04}_{-0.051}$ & 10.014$^{+0.038}_{-0.044}$ & 9.955$^{+0.047}_{-0.042}$ & 9.914$^{+0.043}_{-0.042}$ & 9.905$^{+0.064}_{-0.066}$ & 9.966$^{+0.019}_{-0.018}$ \\
$\log_{10}(M_\star)$ & 1.0 & 9.819$^{+0.075}_{-0.075}$ & 9.955$^{+0.051}_{-0.061}$ & 10.02$^{+0.039}_{-0.048}$ & 9.962$^{+0.071}_{-0.068}$ & 9.928$^{+0.088}_{-0.067}$ & 9.918$^{+0.089}_{-0.098}$ & 9.996$^{+0.027}_{-0.025}$ \\
$\log_{10}(M_\star)$ & 5.0 & 9.887$^{+0.178}_{-0.091}$ & 9.932$^{+0.053}_{-0.066}$ & 9.955$^{+0.073}_{-0.081}$ & 9.765$^{+0.441}_{-0.407}$ & 9.814$^{+1.237}_{-0.391}$ & 9.611$^{+0.821}_{-0.499}$ & 9.978$^{+0.095}_{-0.08}$ \\

	\hline
        $\alpha$ & 0.5 & -1.379$^{+0.036}_{-0.037}$ & -1.38$^{+0.034}_{-0.024}$ & -1.413$^{+0.04}_{-0.034}$ & -1.355$^{+0.05}_{-0.053}$ & -1.271$^{+0.068}_{-0.066}$ & -1.264$^{+0.132}_{-0.107}$ & -1.356$^{+0.018}_{-0.017}$ \\
$\alpha$ & 1.0 & -1.389$^{+0.041}_{-0.041}$ & -1.358$^{+0.053}_{-0.042}$ & -1.425$^{+0.056}_{-0.041}$ & -1.347$^{+0.108}_{-0.098}$ & -1.233$^{+0.166}_{-0.159}$ & -1.293$^{+0.245}_{-0.181}$ & -1.393$^{+0.028}_{-0.028}$ \\
$\alpha$ & 5.0 & -1.291$^{+0.09}_{-0.073}$ & -1.388$^{+0.111}_{-0.085}$ & -1.207$^{+0.139}_{-0.193}$ & -1.111$^{+0.107}_{-1.639}$ & -1.532$^{+0.534}_{-2.203}$ & -1.197$^{+0.18}_{-1.783}$ & -1.365$^{+0.105}_{-0.111}$ \\

	\hline
        $\log_{10}(M_{\rm min})$ & 0.5 & 7.539$^{+0.041}_{-0.052}$ & 7.557$^{+0.038}_{-0.044}$ & 7.626$^{+0.05}_{-0.06}$ & 7.569$^{+0.086}_{-0.095}$ & 7.401$^{+0.149}_{-0.198}$ & 7.356$^{+0.249}_{-0.455}$ & 7.547$^{+0.028}_{-0.029}$ \\
$\log_{10}(M_{\rm min})$ & 1.0 & 7.56$^{+0.068}_{-0.066}$ & 7.544$^{+0.069}_{-0.082}$ & 7.636$^{+0.078}_{-0.096}$ & 7.487$^{+0.18}_{-0.255}$ & 7.169$^{+0.383}_{-0.943}$ & 7.337$^{+0.406}_{-1.189}$ & 7.58$^{+0.042}_{-0.049}$ \\
$\log_{10}(M_{\rm min})$ & 5.0 & 7.306$^{+0.197}_{-0.32}$ & 7.676$^{+0.167}_{-0.246}$ & 6.922$^{+0.641}_{-0.981}$ & 6.476$^{+2.002}_{-0.786}$ & 8.148$^{+0.677}_{-2.121}$ & 7.004$^{+1.416}_{-1.053}$ & 7.502$^{+0.174}_{-0.22}$ \\

	\hline
        $\log_{10}(M_{\rm max})$ & 0.5 & 11.449$^{+1.04}_{-1.124}$ & 11.649$^{+0.855}_{-0.961}$ & 11.825$^{+0.82}_{-0.868}$ & 11.686$^{+0.874}_{-0.915}$ & 11.79$^{+0.809}_{-0.914}$ & 11.803$^{+0.839}_{-0.881}$ & 11.693$^{+0.894}_{-0.87}$ \\
$\log_{10}(M_{\rm max})$ & 1.0 & 11.46$^{+1.066}_{-1.129}$ & 11.666$^{+0.93}_{-0.932}$ & 11.773$^{+0.829}_{-0.8}$ & 11.709$^{+0.831}_{-0.861}$ & 11.712$^{+0.839}_{-0.84}$ & 11.783$^{+0.757}_{-0.852}$ & 11.794$^{+0.829}_{-0.819}$ \\
$\log_{10}(M_{\rm max})$ & 5.0 & 11.234$^{+1.199}_{-1.307}$ & 11.733$^{+0.852}_{-0.965}$ & 11.783$^{+0.867}_{-0.856}$ & 10.866$^{+1.441}_{-1.017}$ & 10.288$^{+1.541}_{-0.49}$ & 10.652$^{+1.433}_{-1.061}$ & 11.747$^{+0.854}_{-0.903}$ \\

	\hline
        $\beta$ & 0.5 & 1.103$^{+1.848}_{-1.51}$ & -0.02$^{+0.984}_{-0.683}$ & 0.002$^{+0.606}_{-0.561}$ & 0.947$^{+0.463}_{-0.529}$ & 1.467$^{+0.416}_{-0.426}$ & 1.418$^{+0.602}_{-0.614}$ & 0.972$^{+0.039}_{-0.042}$ \\
$\beta$ & 1.0 & 1.142$^{+1.816}_{-1.493}$ & 0.591$^{+1.258}_{-1.019}$ & -0.173$^{+0.733}_{-0.579}$ & 0.75$^{+0.859}_{-0.892}$ & 1.346$^{+0.73}_{-0.951}$ & 1.098$^{+0.999}_{-0.938}$ & 0.968$^{+0.042}_{-0.039}$ \\
$\beta$ & 5.0 & 1.596$^{+1.62}_{-1.693}$ & 1.013$^{+1.792}_{-1.395}$ & 1.163$^{+1.377}_{-1.45}$ & 2.333$^{+1.11}_{-1.776}$ & 1.876$^{+1.42}_{-1.698}$ & 1.966$^{+1.268}_{-1.519}$ & 0.967$^{+0.042}_{-0.04}$ \\

	\hline
	\hline
	$\Omega_{\rm HI} \times 10^4 \;(\rm individual)$ & 0.5 & 3.76$^{+1.13}_{-0.89}$ & 4.79$^{+0.67}_{-0.3}$ & 5.7$^{+0.24}_{-0.35}$ & 6.02$^{+0.36}_{-0.24}$ & 6.63$^{+0.28}_{-0.35}$ & 6.71$^{+0.58}_{-0.44}$ &  \\
$\Omega_{\rm HI} \times 10^4 \;(\rm individual)$ & 1.0 & 3.68$^{+1.08}_{-0.81}$ & 5.0$^{+0.54}_{-0.4}$ & 5.58$^{+0.41}_{-0.31}$ & 5.85$^{+0.43}_{-0.37}$ & 6.29$^{+0.6}_{-0.52}$ & 6.39$^{+0.98}_{-0.81}$ &  \\
$\Omega_{\rm HI} \times 10^4 \;(\rm individual)$ & 5.0 & 4.05$^{+1.1}_{-1.29}$ & 5.24$^{+0.65}_{-0.59}$ & 5.28$^{+1.09}_{-0.87}$ & 4.94$^{+1.35}_{-1.38}$ & 8.12$^{+2.13}_{-2.09}$ & 6.07$^{+3.38}_{-3.84}$ &  \\

	\hline
	$\Omega_{\rm HI} \times 10^4 \;(\rm broad)$ & 0.5 & 4.84$^{+0.17}_{-0.16}$ & 5.21$^{+0.16}_{-0.16}$ & 5.62$^{+0.16}_{-0.16}$ & 6.05$^{+0.16}_{-0.16}$ & 6.47$^{+0.17}_{-0.16}$ & 6.78$^{+0.19}_{-0.17}$ &  \\
$\Omega_{\rm HI} \times 10^4 \;(\rm broad)$ & 1.0 & 4.72$^{+0.2}_{-0.17}$ & 5.09$^{+0.19}_{-0.18}$ & 5.5$^{+0.2}_{-0.18}$ & 5.91$^{+0.2}_{-0.18}$ & 6.34$^{+0.21}_{-0.2}$ & 6.64$^{+0.22}_{-0.22}$ &  \\
$\Omega_{\rm HI} \times 10^4 \;(\rm broad)$ & 5.0 & 4.64$^{+0.4}_{-0.48}$ & 5.0$^{+0.43}_{-0.5}$ & 5.4$^{+0.47}_{-0.53}$ & 5.8$^{+0.5}_{-0.57}$ & 6.21$^{+0.54}_{-0.61}$ & 6.5$^{+0.57}_{-0.64}$ &  \\

	\hline
	\end{tabular}
\end{table*}

\subsection{The reconstructed HIMF from our baseline survey}\label{sec:baselineresults}

In the HIMF parameterisation, the combined $\alpha$ and $\phi_\star$  parameters determine the slope and normalisation of the low mass end of the HIMF, while $M_{\rm min}$ provides the \ha mass at which the HIMF can be measured down to. The combination of $M_\star$ and $\phi_\star$ determines where the knee of the mass function occurs. The $M_{\rm max}$ parameter indicates the upper mass limit that can be constrained by the data. In all cases considered here $M_{\rm max}$ is not well constrained, but this has very little effect on the fits due to the exponential cut-off at high masses in the Schechter function always giving very few galaxies beyond $M_\star$. 
The parameter defining redshift dependence, $\beta$, is sensitive to the number of sources in individual redshift bins.

Fig.~\ref{fig:hist_himf} shows the source counts for the measured \ha flux $S_m$ alongside the reconstructed HIMFs model for our baseline survey. 
 The green line shows the results recovered from individual redshift bins, where we fix $\log_{10}(\phi_{\star}) = -2.318$ and then constrain the other 5 parameters. This breaks the degeneracy between $\phi_\star$ and $\beta$ in a given individual redshift bin. Since we are assuming a pure density evolution of the HIMF $\phi_\star$ and $\beta$ are completely degenerate in the sense one can compensate for a high $\phi_\star$ with a low $\beta$ and vice versa. The blue line in Fig.~\ref{fig:hist_himf} shows the results recovered using the evolving HIMF model in a single broad redshift bin spanning $0<z<0.55$. As shown, the model distributions of $S_m$ and the resulting HIMF, using both methods, are in good agreement with the input simulated data across all redshift bins.

\begin{figure*}
  \centering
    \begin{subfigure}[b]{0.5\textwidth}
    \includegraphics[width=1.06\columnwidth]{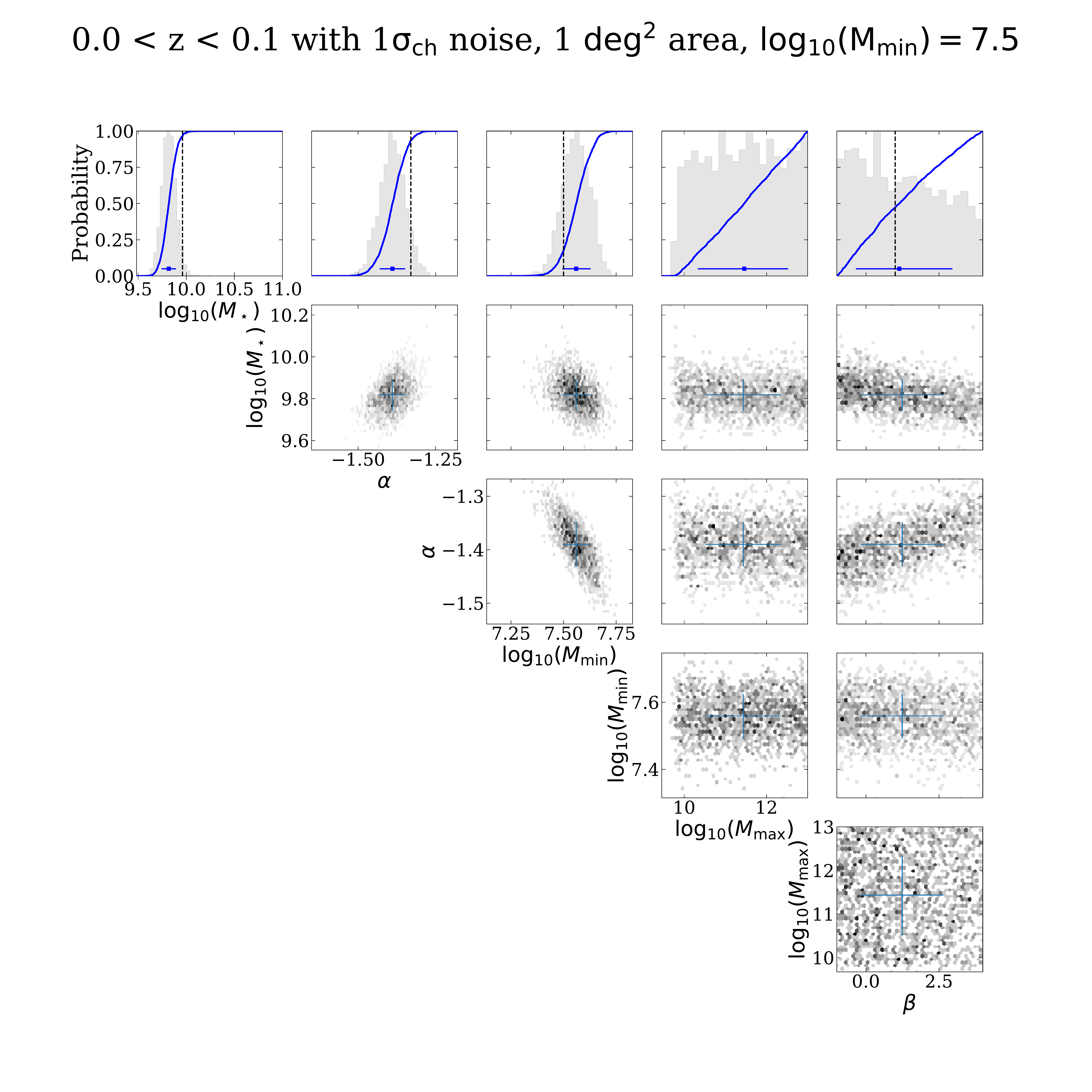}
  \end{subfigure}%
  \hfill
  \begin{subfigure}[b]{0.5\textwidth}
    \includegraphics[width=1.06\columnwidth]{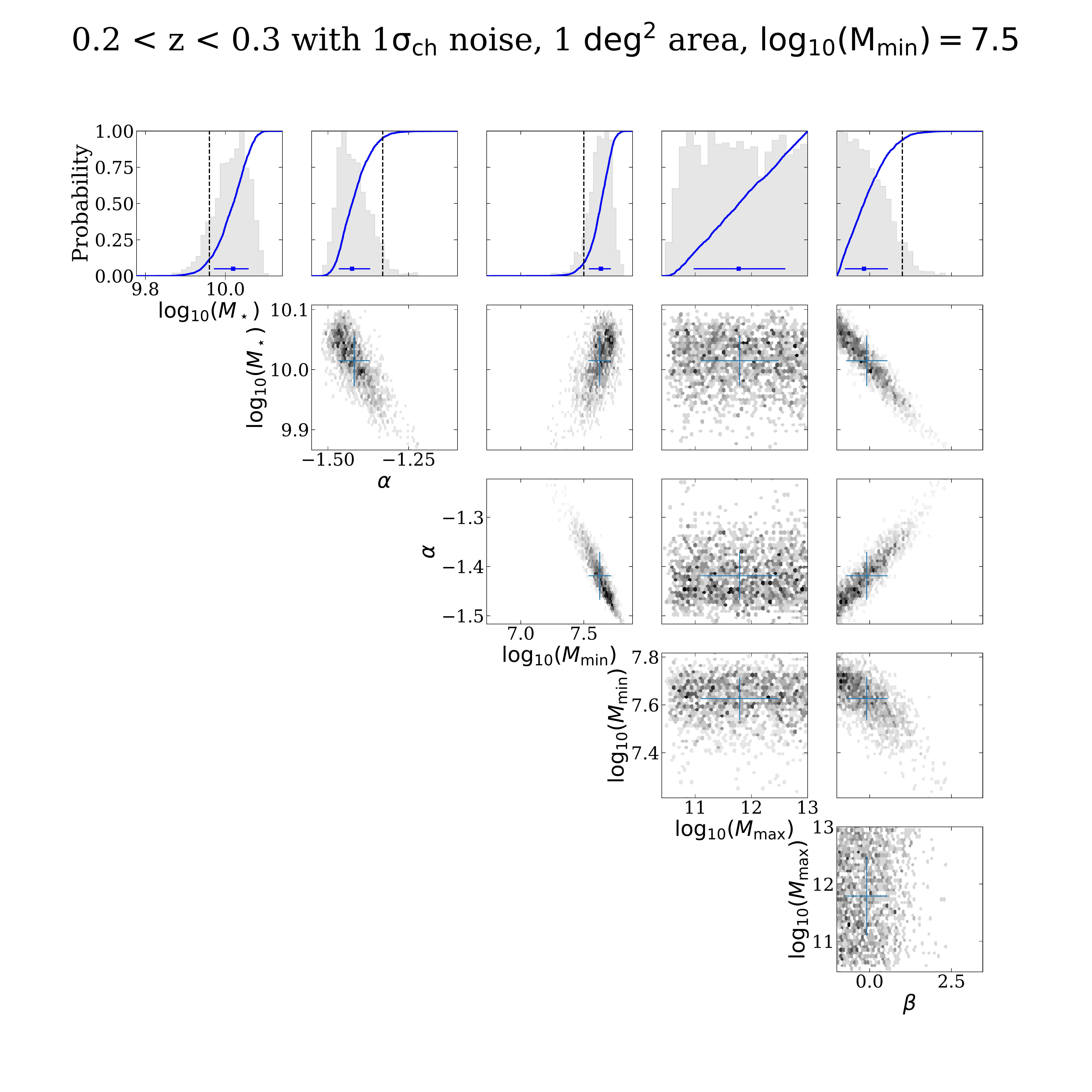}
  \end{subfigure}%
    \hfill
    \begin{subfigure}[b]{0.5\textwidth}
    \includegraphics[width=1.06\columnwidth]{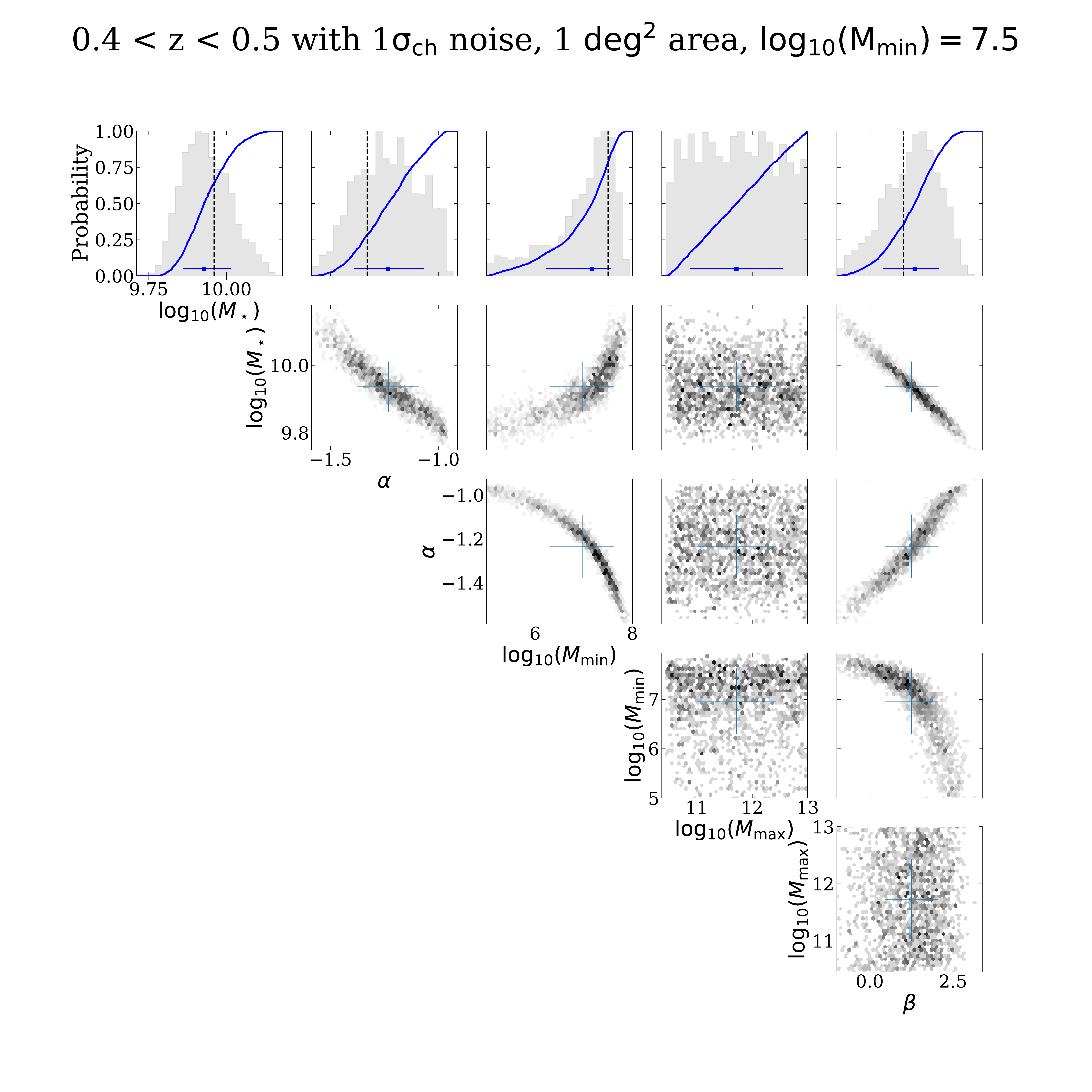}
  \end{subfigure}%
  \hfill
  \raggedleft
  \begin{subfigure}[b]{0.5\textwidth}
    \includegraphics[width=1.1\columnwidth, height=1.06\columnwidth]{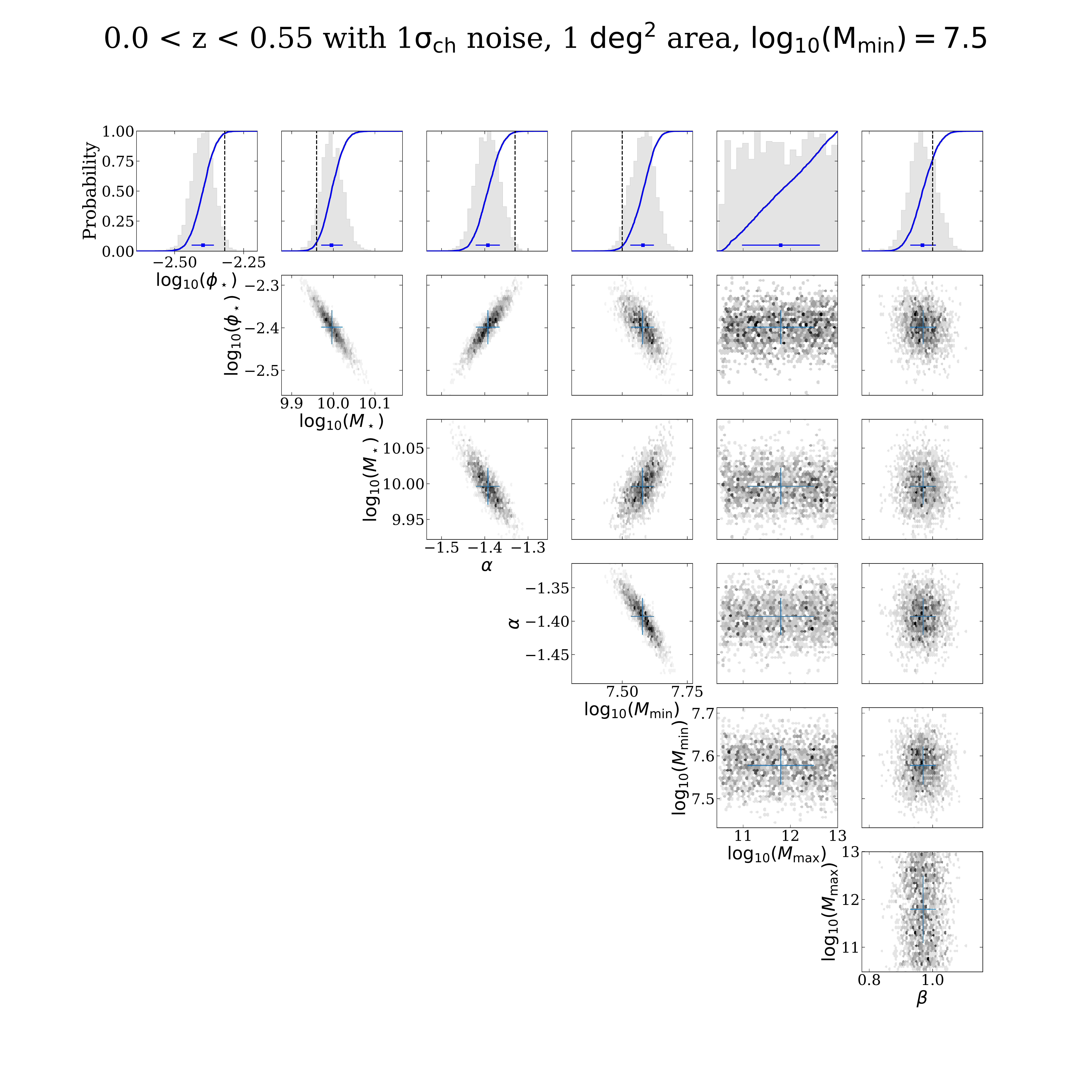}
  \end{subfigure}%

  \caption{The posterior distributions for HIMF model parameters for three individual redshift bins for our baseline survey (top and bottom left panels) and for the single broad redshift bin (bottom right panel). The grey histograms are the (1 or 2\,dimensional) marginal posterior probability densities. The blue curves are the cumulative distribution (integrating over the grey histograms from left to right). The best-fitting parameters are shown by the blue dots with 1\,$\sigma$ error bars, while the input parameter values are the vertical black lines. Please refer to Fig.~\ref{fig:marg_m} for bigger labels.}
  \label{fig:marg_zbin}
\end{figure*}

Taking a closer look at the recovered HIMF from the individual redshift bins, in Fig.~\ref{fig:marg_zbin} we show the posterior probability distributions for the parameters of the HIMF model for 3 bins in redshift and our fits to the parameters are given in Table~\ref{tab:pos_s}. We find that the constraints on the low and high mass ends of the HIMF evolve differently. The signal-to-noise is significantly reduced at higher redshifts due to increasing luminosity distance and this leads to a significantly weaker constraint on the low-mass end with increasing redshift. However, for the high-mass end, the sources are sufficiently bright that the noise does not significantly affect the measured line fluxes out to intermediate ($z\sim 0.3)$ redshift. In this regime, as the number of bright sources increases with the increasing volume they work to strengthen the constraints on the position of the knee of the mass function. However, eventually this effect is offset by the reduction in signal-to-noise due to the increasing luminosity distance at $z>0.3$.  We therefore find the tightest constraints on $M_{\star}$ in the middle of the redshift range considered, with other redshift bins biased to low values of $M_\star$, due to the lack of volume to sample the most massive galaxies at low redshifts and the strong effect of the noise on the faint end slope at high redshifts for this particular simulation. This trend is also apparent for the average \ha mass density $\Omega_{\rm HI}$(individual), as shown in the bottom rows of Table~\ref{tab:pos_s}, where the accuracy of $\Omega_{\rm HI}$ is very much tied to the accuracy of $M_{\star}$ as this is where the bulk of the mass density resides. The evolution parameter is not well constrained from individual redshift bins in general, due to the limited number of sources in each bin. 

One method of overcoming the deficiencies highlighted by considering smaller individual redshift slices is to model the evolution across the full range in redshift with a single model (blue line/region in Fig.~\ref{fig:hist_himf}). Fig.~\ref{fig:marg_zbin} (bottom right panel) shows the posterior probability distributions for the parameters in the broad redshift bin and the last column of Table~\ref{tab:pos_s} gives a summary of the parameter fits. We see in this case that the fits are much closer to the input values. This is because the faint end slope is highly constrained in the lowest redshift bins and the non-evolving value of $M_\star$ is well constrained at the intermediate redshifts. The high-mass end then just follows the fixed exponential fall off. Given that these parameters are highly constrained, then the evolution parameter is also well constrained and within the uncertainties of the input value. Obviously, this is an idealised case, but does show where having a large observed volume can help constrain the mass function at low redshift, under the assumption that the fundamental shape does not change. Alternatively, we could incorporate the evolutionary term into the $M_\star$ parameter and adopt a pure-mass evolution model. In this case we could then use the Bayesian evidence to distinguish between pure mass and pure density evolution, and we leave this to future work using real data.

The overall benefits of using a broad redshift bin can also be appreciated by considering the constraints on $\Omega_{\rm HI}$(broad), where there is much closer agreement between the input and derived values.

\subsection{The effect of changing the background noise} 

\begin{figure*}
  \centering

  \begin{subfigure}[b]{0.5\textwidth}
    \includegraphics[width=\columnwidth]{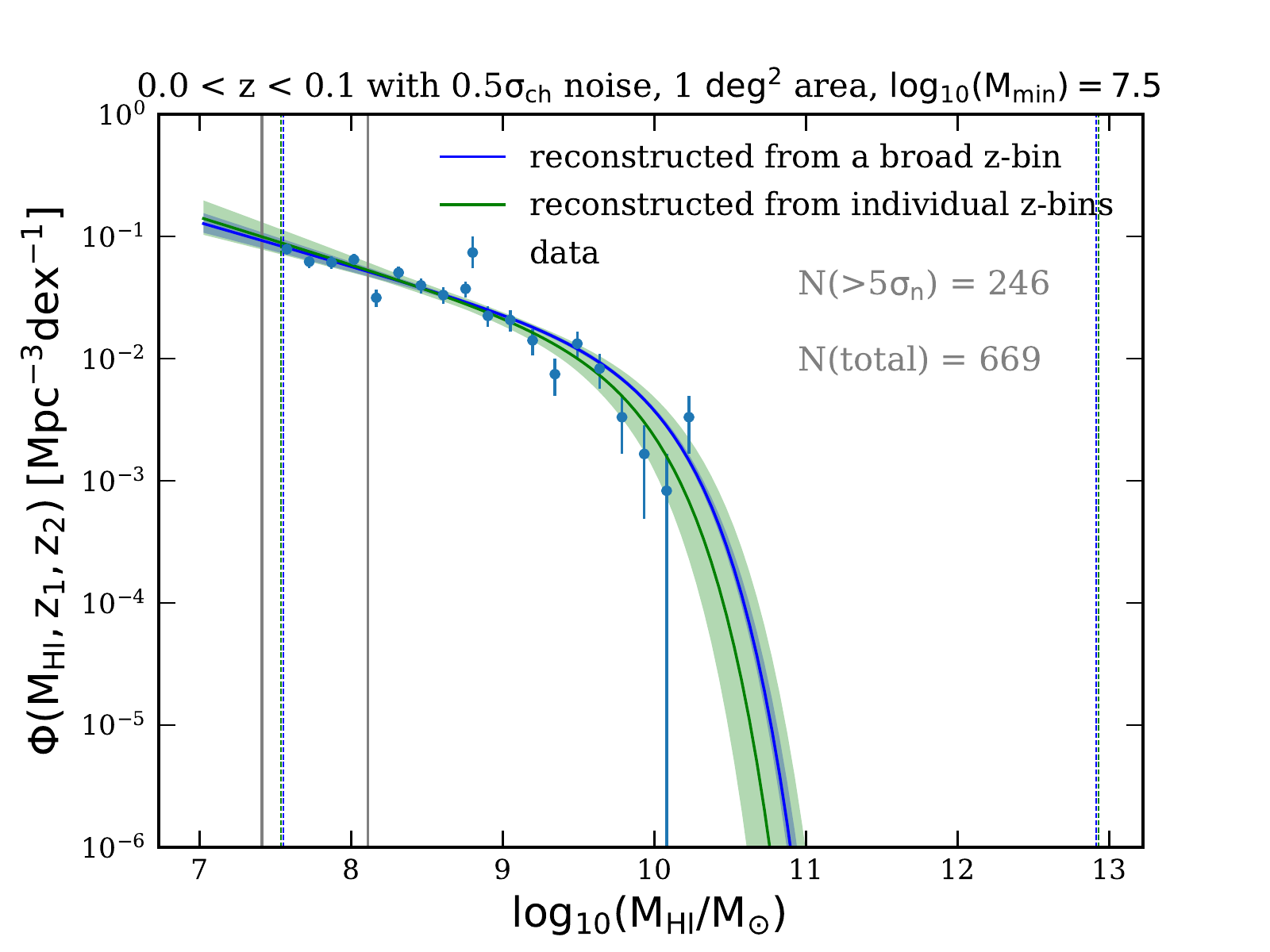}
    \includegraphics[width=\columnwidth]{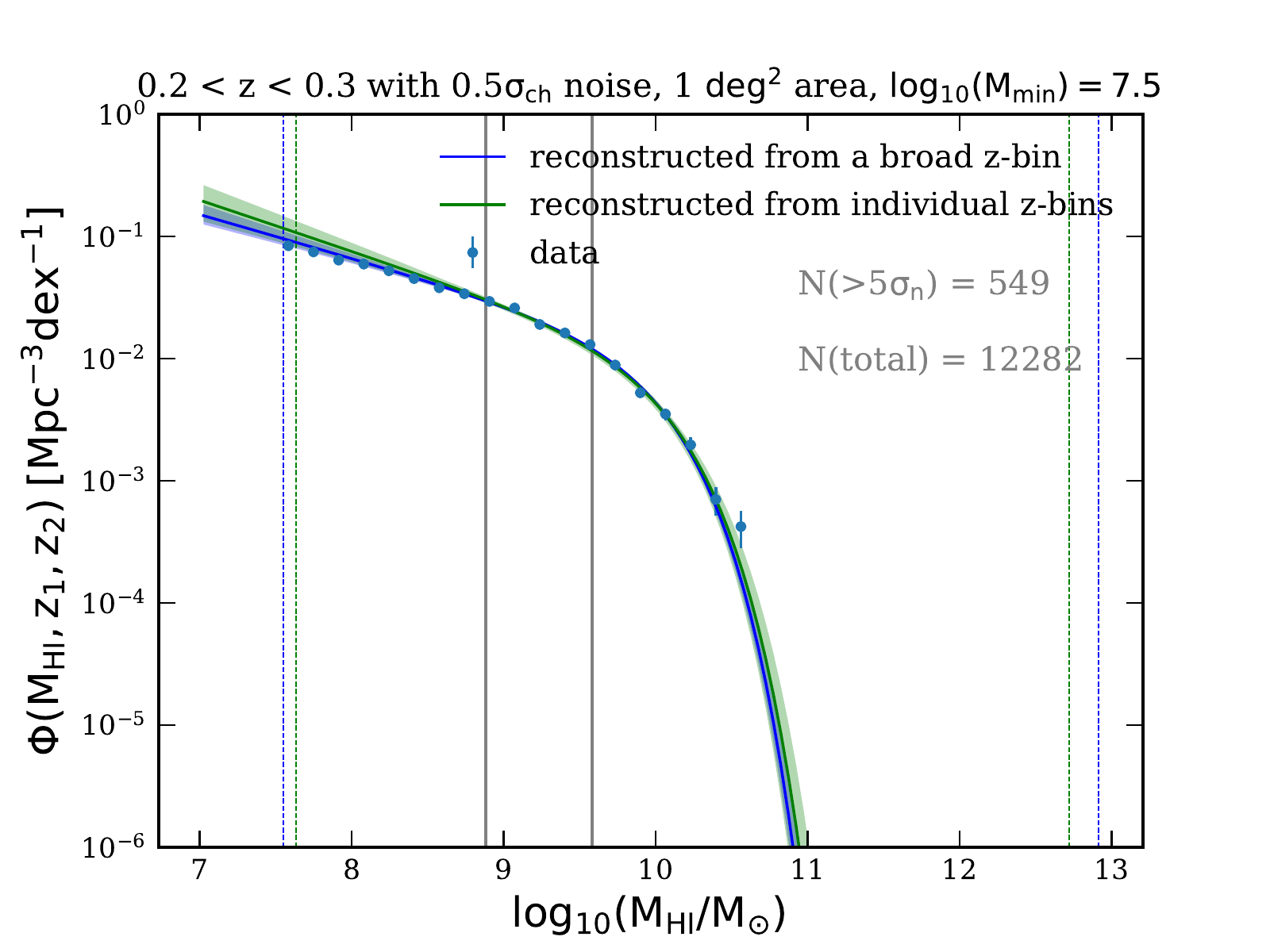}
    \includegraphics[width=\columnwidth]{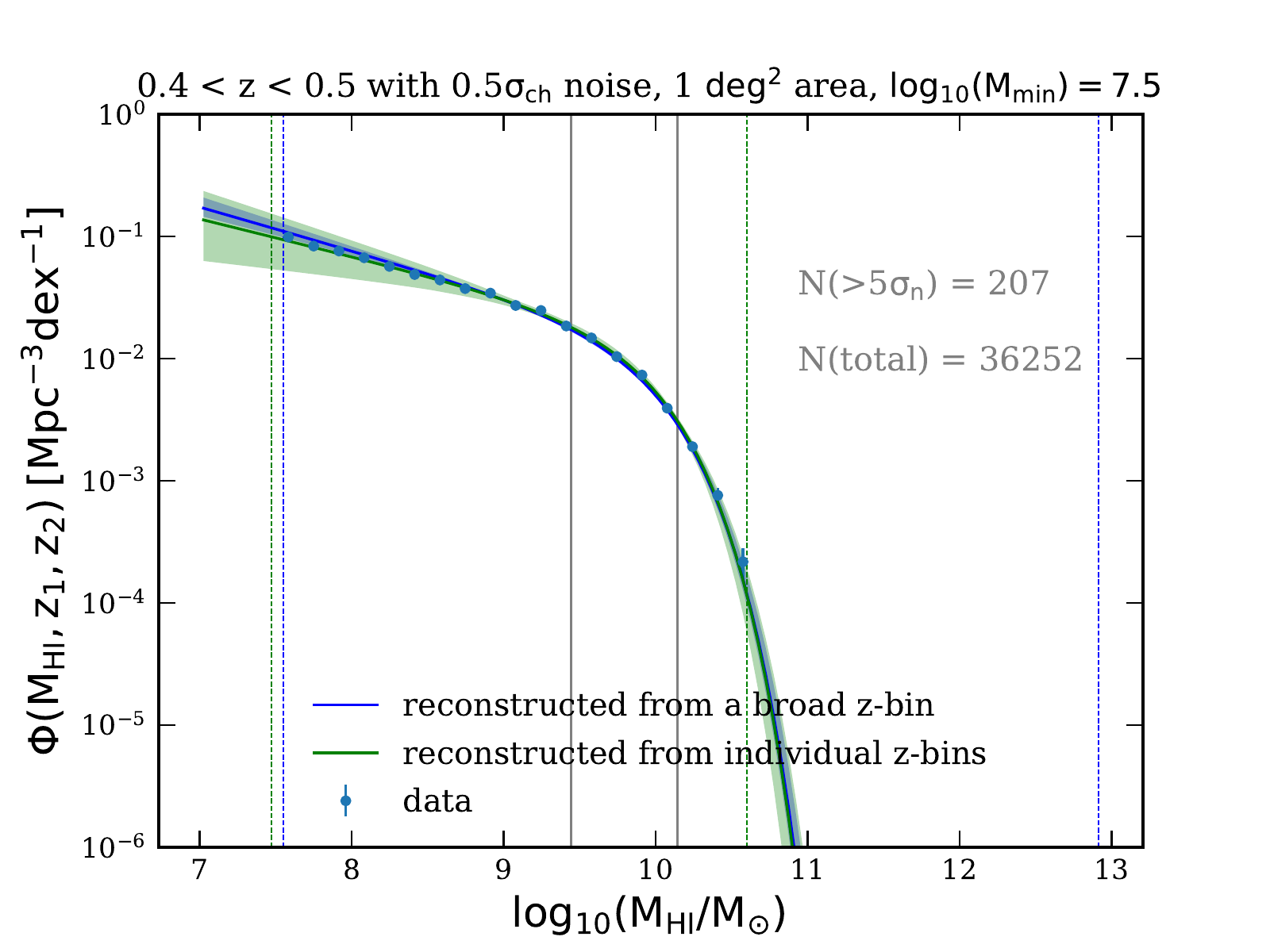}
  \end{subfigure}%
  \hfill
  \begin{subfigure}[b]{0.5\textwidth}
    \includegraphics[width=\columnwidth]{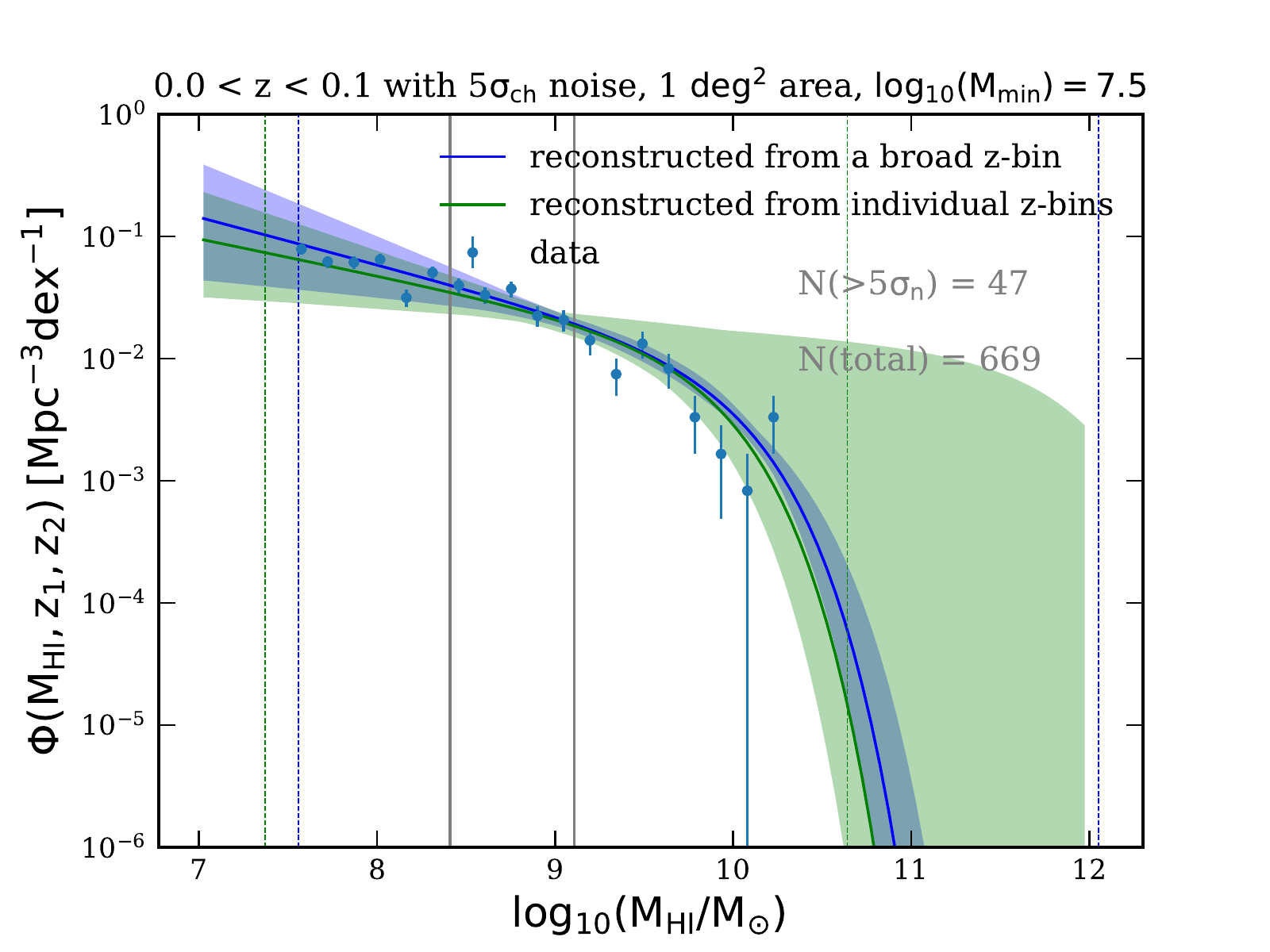}
    \includegraphics[width=\columnwidth]{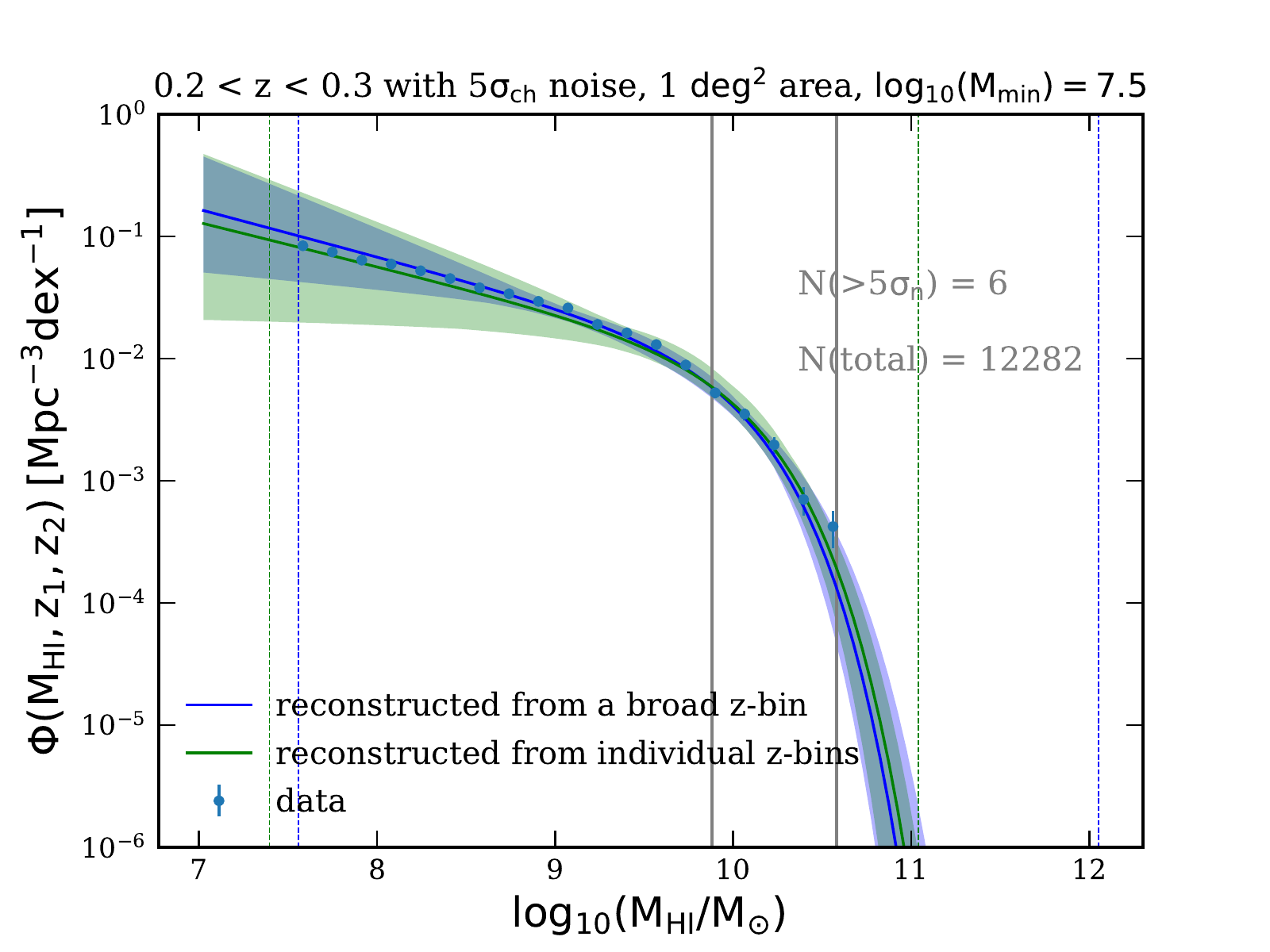}
    \includegraphics[width=\columnwidth]{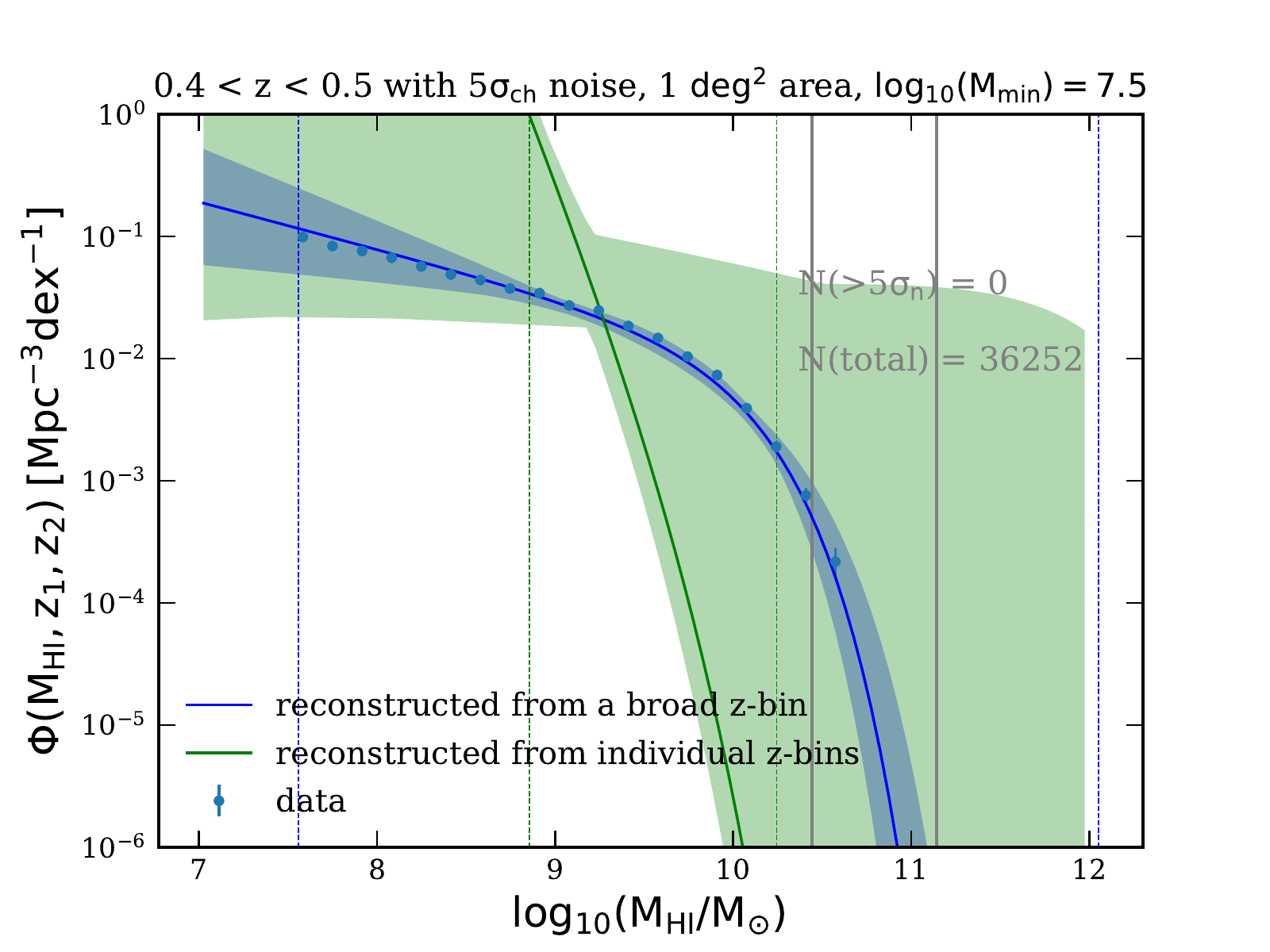}
  \end{subfigure}%
    \caption{Reconstructed HIMF from our simulated survey with 0.5\,$\sigma_{\rm ch}$ (left) and 5\,$\sigma_{\rm ch}$ (right) Gaussian background noises in different redshift bins. The blue lines show results for a single broad redshift bin ( i.e.  $0 < z < 0.55$), while the green lines show the results for individual redshift bins. The color-coded regions are the 68\% credible intervals in the HIMF estimated from the posterior samples. The color-coded vertical lines are $\log_{10}(M_{\rm min})$ and $\log_{10}(M_{\rm max})$ of the best fitting from a broad redshift bin and an individual redshift bin. The grey vertical lines on the left panel indicate the $\sigma_{\rm n} = \sqrt{N_{\rm ch}} \times 0.5 \sigma_{\rm ch} d\vi$ and the formal 5\,$\sigma_{\rm n}$ detection threshold at the centre of each redshift bin, while those on the right panel indicate the $\sigma_{\rm n} = \sqrt{N_{\rm ch}} \times 5\,\sigma_{\rm ch} d\vi$ and 5\,$\sigma_{
    \rm n}$ detection threshold. The errors on the simulated data points are just Possion errors. Only three redshift bins are shown for simplicity.}
    \label{fig:himf_s}
\end{figure*}

\begin{figure*}
 \centering
 \begin{subfigure}[b]{0.5\textwidth}
    \includegraphics[width=\columnwidth, height=\columnwidth]{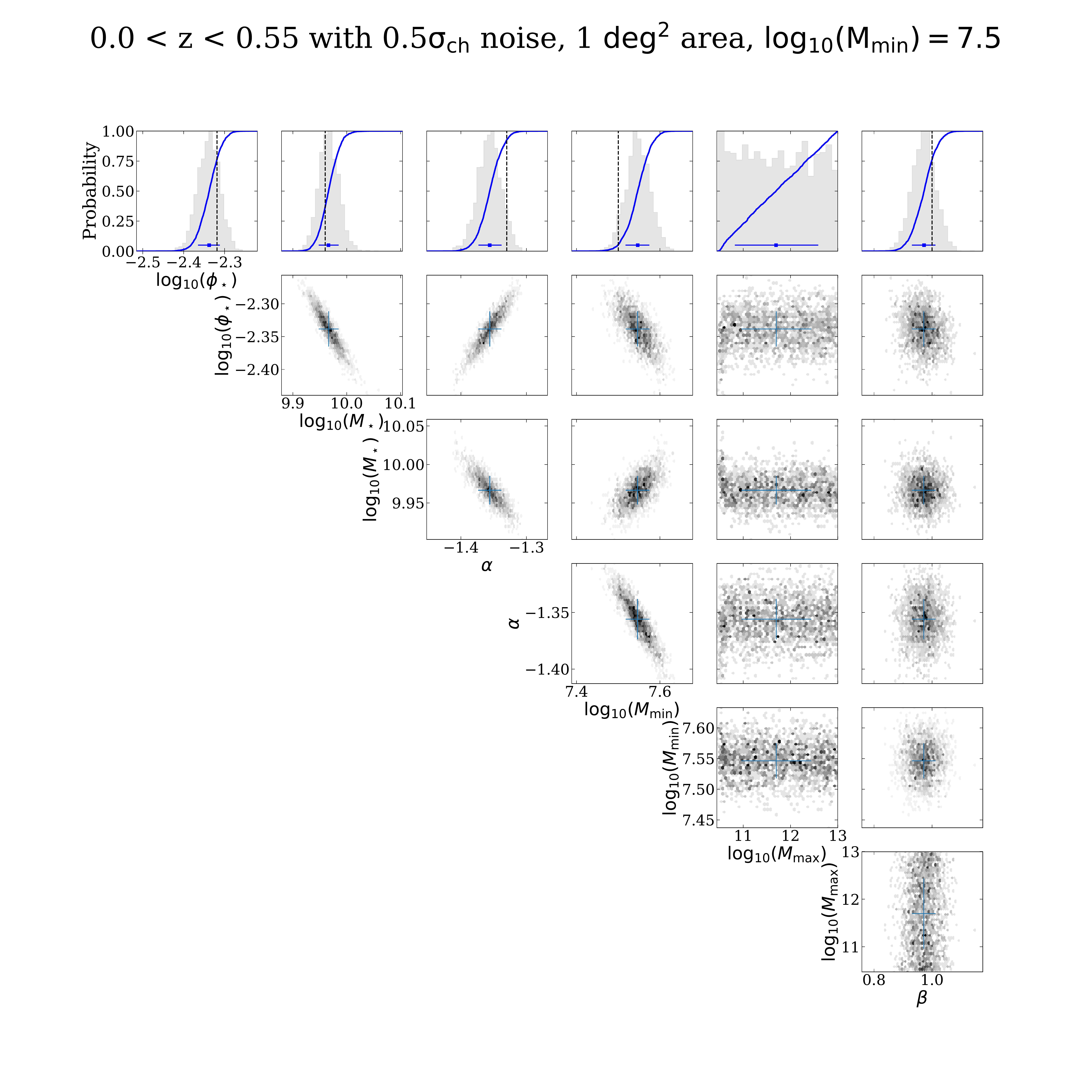}
 \end{subfigure}%
 \hfill
 \begin{subfigure}[b]{0.5\textwidth}
    \includegraphics[width=\columnwidth, height=\columnwidth]{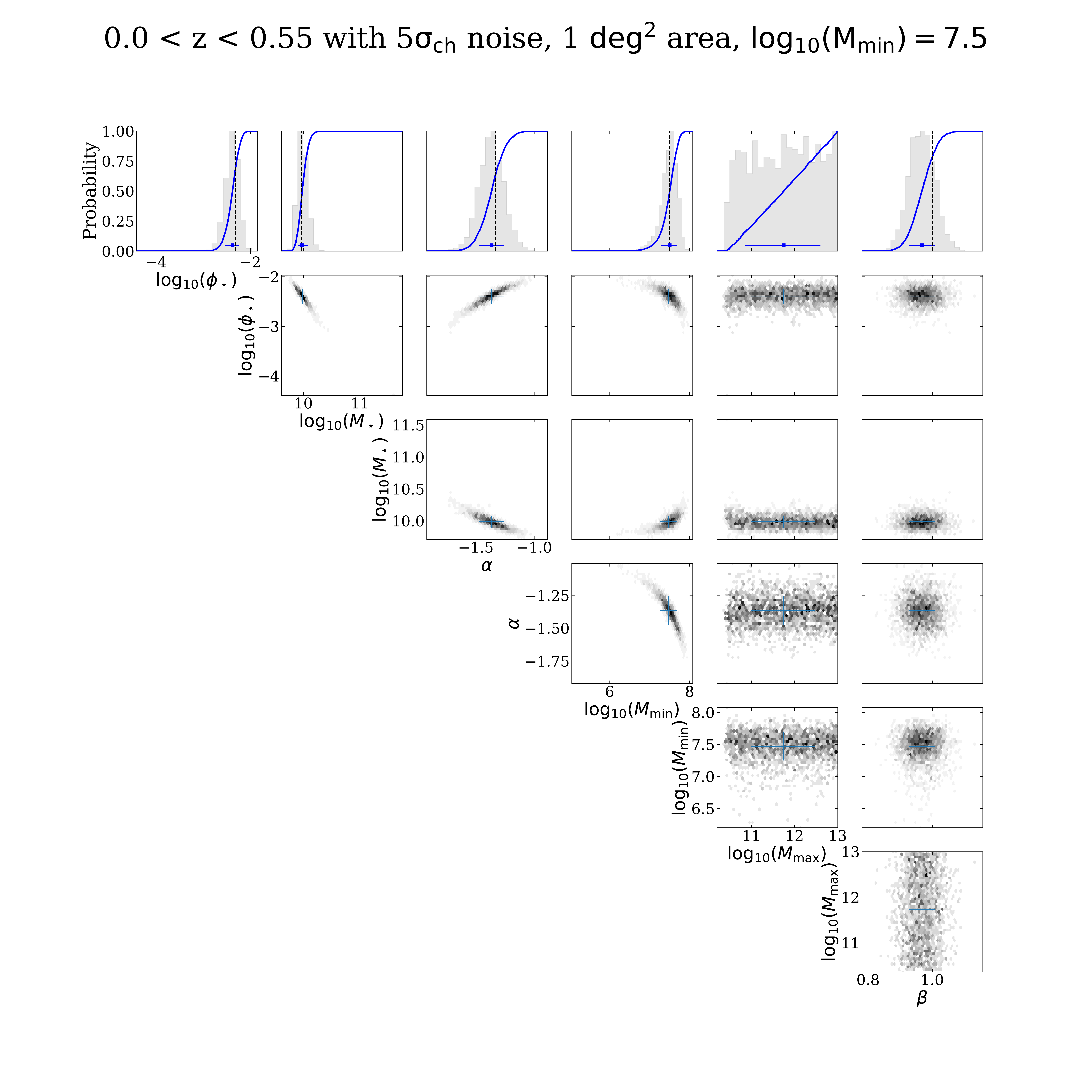}
 \end{subfigure}%
  \caption{The posterior distributions for the HIMF model parameters, for a single broad redshift bin with 0.5\,$\sigma_{\rm ch}$ (left) and 5\,$\sigma_{\rm ch}$ (right) background noise levels. 
  All other properties of this figure are the same as this in Fig.~\ref{fig:marg_zbin}. }
  \label{fig:marg_s}
\end{figure*}

We now look at the effect of changing the noise on the reconstructed HIMF from our baseline survey. We show the resulting parameters for 0.5 and 5$\sigma_{\rm ch}$ noise in Table~\ref{tab:pos_s} for the individual redshift bins and the whole redshift range, respectively. Fig.~\ref{fig:himf_s} shows the HIMF, and its evolution, reconstructed from the simulated data for these different levels of noise. We only show the relevant figures for the model fit over the full redshift range in Fig.~\ref{fig:marg_s}, rather than for the individual bins as the general trends remain the same as discussed in Sec.~\ref{sec:baselineresults}. A subtle difference is that the tightest constraints on $M_{\star}$ and $\Omega_{\rm HI}$ for the 0.5 $\sigma_{\rm ch}$ noise has drifted away from the middle of the redshift range towards higher range considered due to the increased signal-to-noise for all sources.

As expected, increasing the noise by a factor of five leads to significant increase in the uncertainties for the parameters which define the Schechter function and determine $\Omega_{\rm HI}$, particularly for the single redshift bin. The absolute values for the Schechter function parameters (Table~\ref{tab:pos_s}) show that increasing the noise leads to $\phi_\star$ increasing and then being compensated for with a flatter faint end slope and a lower $M_\star$. Again this can be understood by considering the redshift range where the tightest constraints are coming from for this noisier data set. The constraints on the low-mass slope are most affected as our ability to correctly assign the correct flux to the low-mass sources is significantly inhibited by the increased noise in all redshift bins. This in turn leads to a more significant degeneracy between $M_\star$ and $\phi_\star$, where the 2-dimensional posterior distribution changes from a linear degeneracy towards a more "banana-shape" (Fig.~\ref{fig:marg_s}).

On the other hand, the values for the $\beta$ are not affected significantly by increasing the noise, due to the fact that once the Schechter function parameters are set in place then the evolution term mainly depends on the number of sources in the most constrained mass bins at all redshifts, which are generally slightly lower than $M_\star$ and therefore are not significantly degenerate with $M_\star$ or $\alpha$.

\subsection{The effect of increasing the minimum mass of the \ha galaxy sample}

Any parent spectroscopic redshift catalogue for galaxies used to conduct an \ha stacking experiment will inevitably be flux limited. This naturally leads to a limit on the \ha mass that any model function can be fit down to. Furthermore, this may also change over the redshift range of interest. In this section, we therefore investigate this by simply increasing the minimum mass from $M_{\rm min} = 10^{7.5}$ to $10^{8.5}$\,M$_{\odot}$ on the reconstructed HIMF. 

\begin{figure*}
  \centering
  \begin{subfigure}[b]{0.33\textwidth}
    \includegraphics[width=1.1\columnwidth, height=0.86\columnwidth]{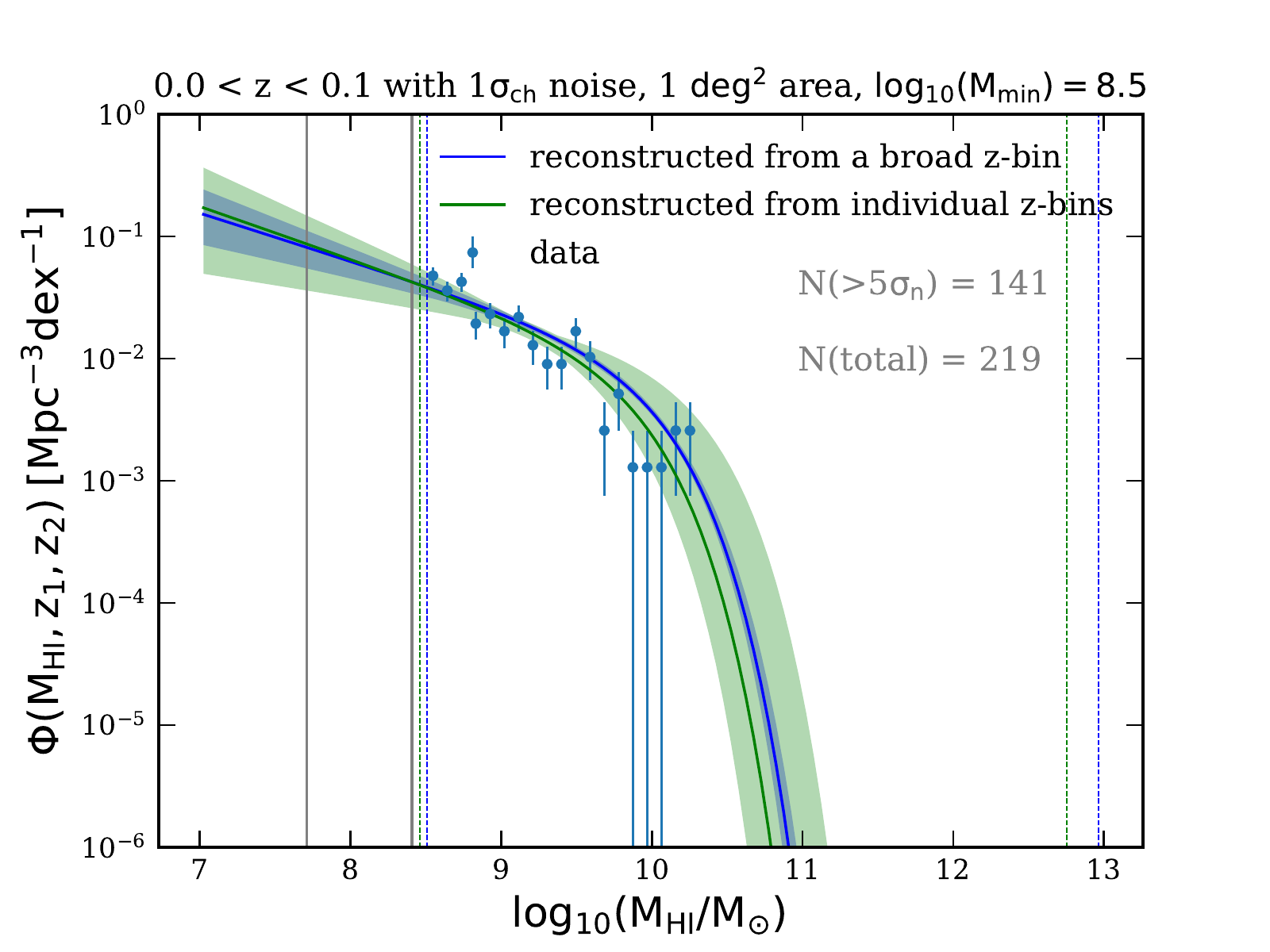}
  \end{subfigure}%
  \hfill
  \begin{subfigure}[b]{0.33\textwidth}
    \includegraphics[width=1.1\columnwidth, height=0.86\columnwidth]{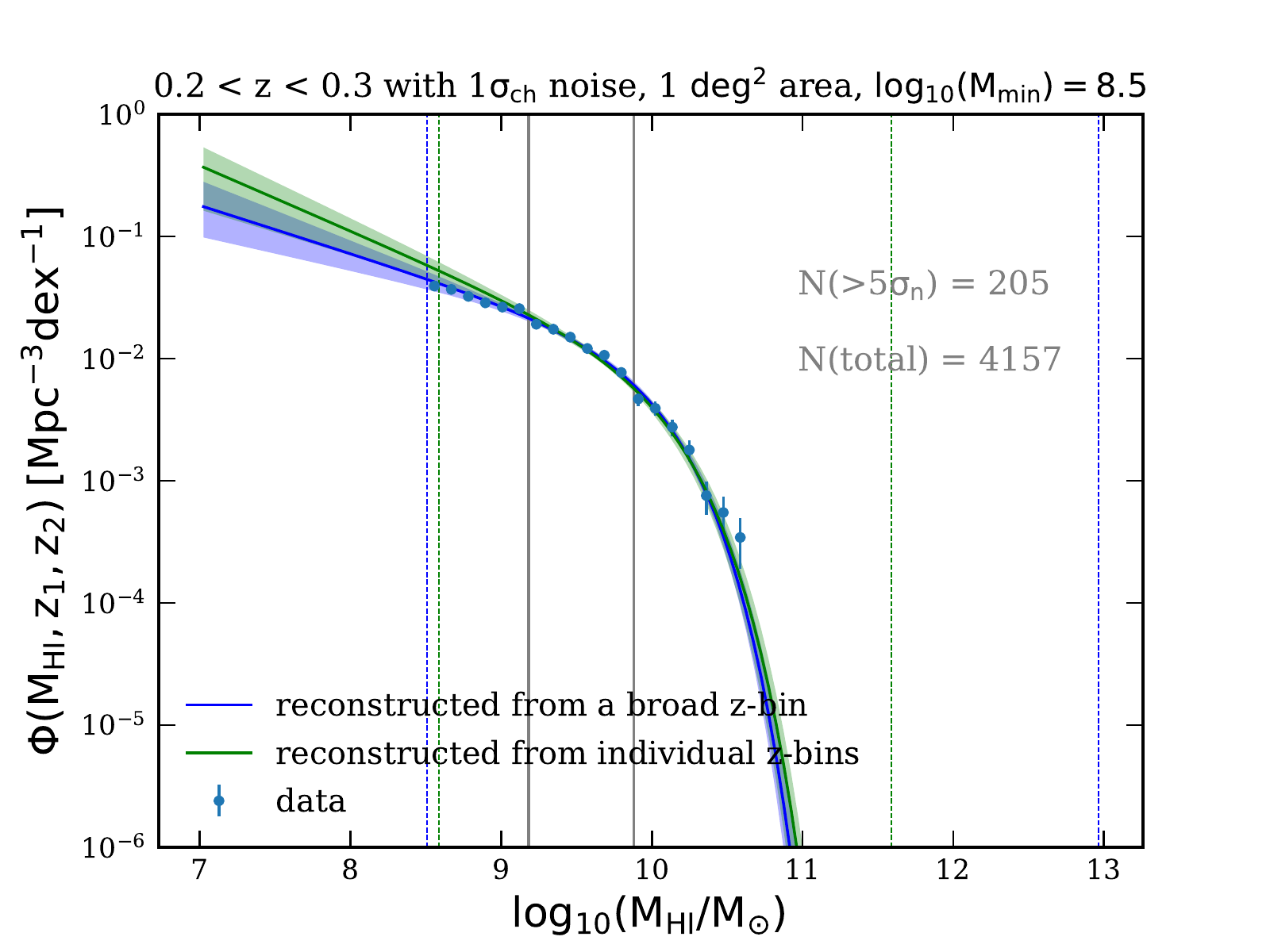}
  \end{subfigure}%
  \hfill
  \begin{subfigure}[b]{0.33\textwidth}
    \includegraphics[width=1.1\columnwidth, height=0.86\columnwidth]{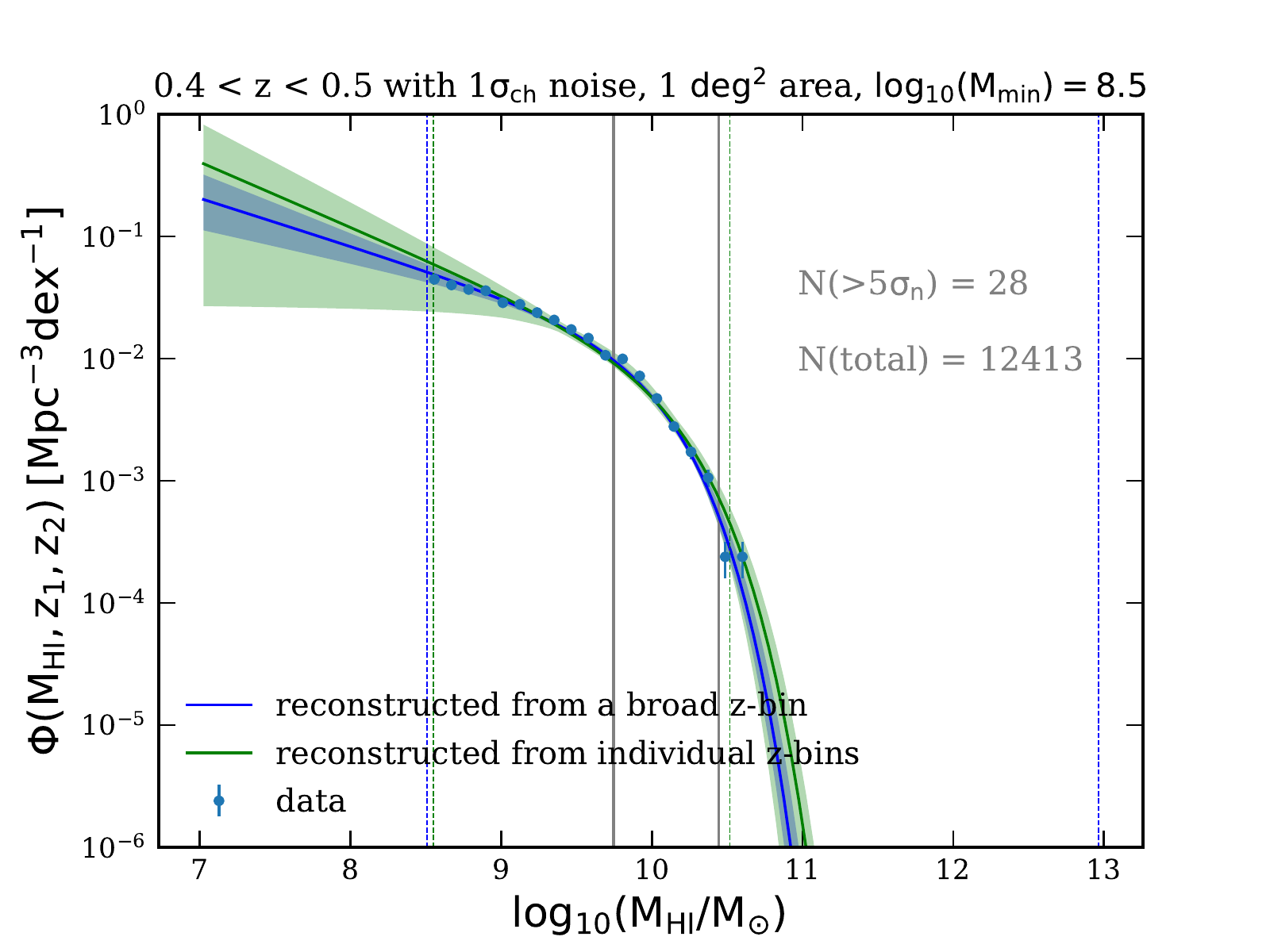}
  \end{subfigure}%
    \caption{Reconstructed HIMF from our simulated survey with increasing minimum mass  $\log_{10}(M_{\rm min}) = 8.5$ of the \ha galaxy sample in different redshift bins. The blue lines show the results from a single broad redshift bin ( i.e.  $0 < z < 0.55$) while the green lines show the results for individual redshift bins. The grey vertical lines indicate the $\sigma_{\rm n} = \sqrt{N_{\rm ch}} \sigma_{\rm ch} d\vi$ and 5\,$\sigma_{\rm n}$ detection threshold at the centre of each redshift bin. All other properties of this figure are the same as that of Fig.~\ref{fig:himf_s}.}
    \label{fig:himf_m}
\end{figure*}

\begin{figure*}
    \includegraphics[width=\textwidth]{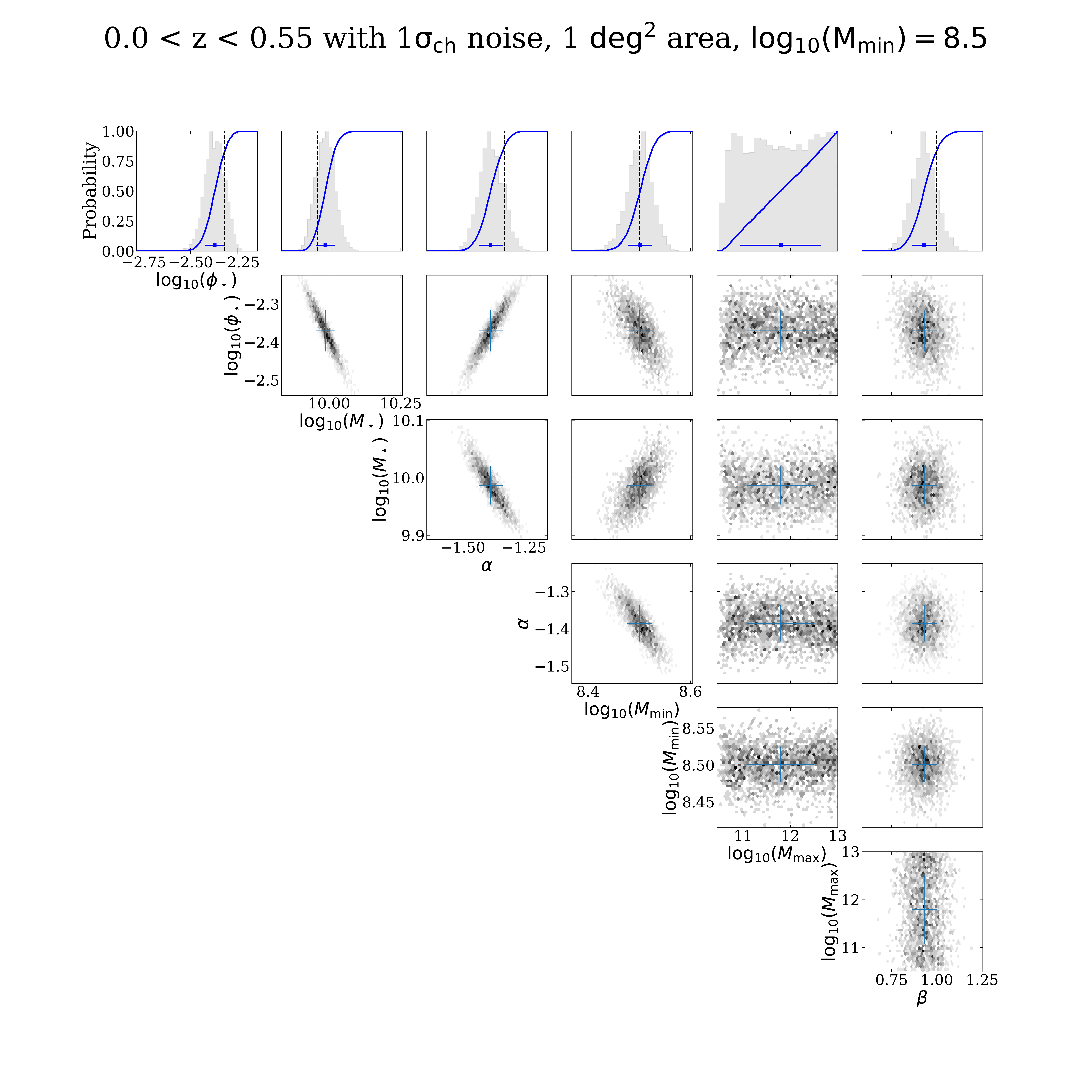}
 \caption{The posterior distributions for the HIMF model parameters for a single broad redshift bin from our simulated survey with increasing minimum mass  $\log_{10}(M_{\rm min}) = 8.5$ of the \ha galaxy sample. All other properties of this figure are the same as that of Fig.~\ref{fig:marg_zbin}. }
 \label{fig:marg_m}
\end{figure*}

In Fig.~\ref{fig:himf_m} we show the resulting HIMF model and its evolution with redshift, with the input catalogue limited to  $M_{\rm HI} > 10^{8.5}$\,M$_{\odot}$. Posterior values for the parameters are given in Table~\ref{tab:pos_m} and Fig.~\ref{fig:marg_m}. As before we do not present the posterior distributions for the individual redshift bins.

Obviously, imposing a higher lower-mass limit to the \ha sample will automatically lead to a smaller sample overall and naturally lead to poorer constraints on the low-mass slope of the HIMF and we see this in our results. However, the constraints on $M_\star$ and $\phi_\star$ are not affected as much. Reassuringly, our fit does find where the HIMF model is reliably fit to, where we find $M_{\rm min}$ increases from $10^{7.5}$ to $10^{8.5}$\,M$_{\odot}$ when we alter the lower mass limit of the sample. The $\Omega_{\rm HI}$ is less affected at higher redshifts since $M_\star$ and $\phi_\star$ are still well constrained.

\subsection{The effect of changing the survey area } 

\begin{figure*}
  \centering    
    \begin{subfigure}[b]{0.5\textwidth}
    \includegraphics[width=1.06\columnwidth]{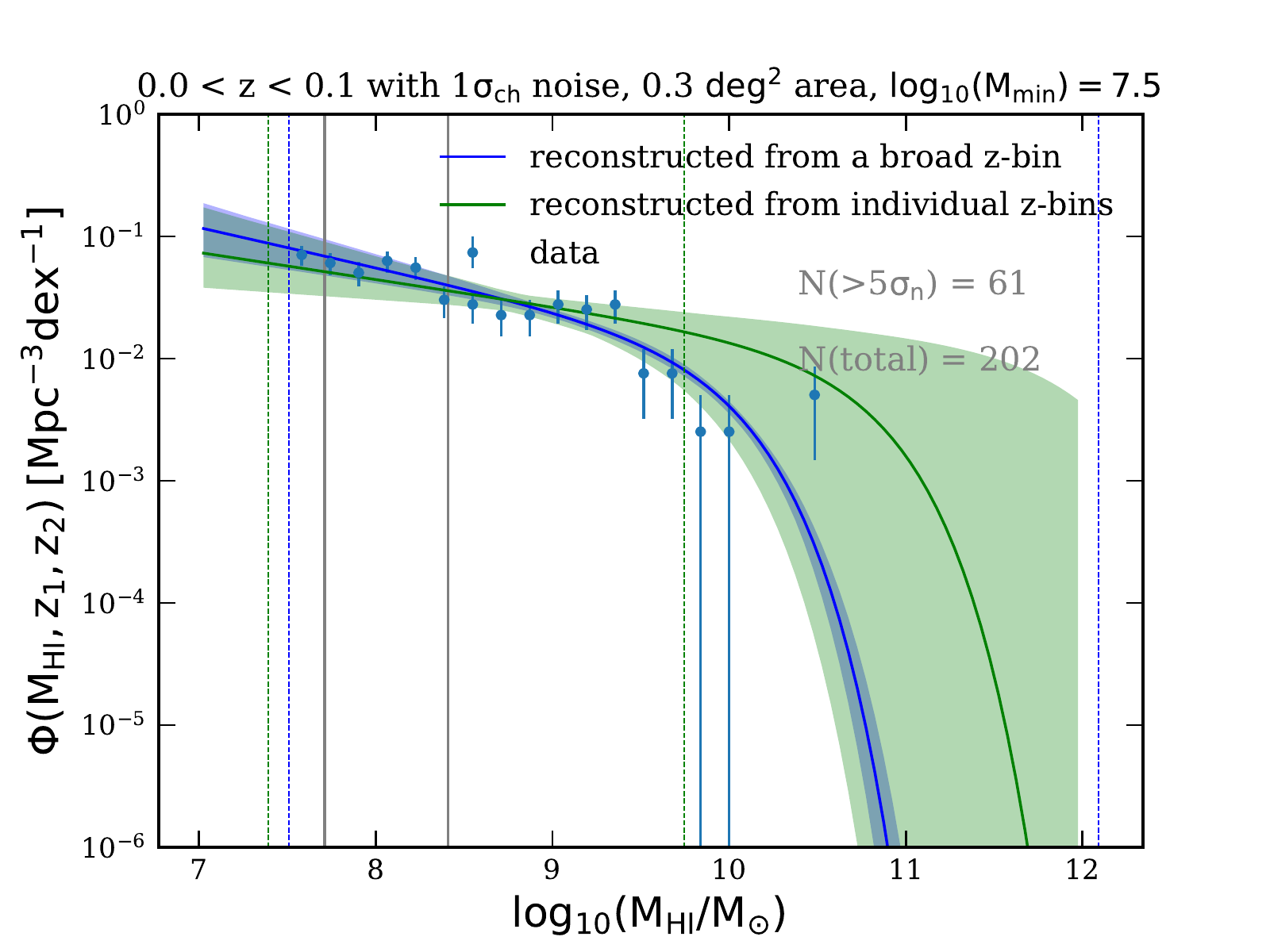}
  \end{subfigure}%
  \hfill
  \begin{subfigure}[b]{0.5\textwidth}
    \includegraphics[width=1.06\columnwidth]{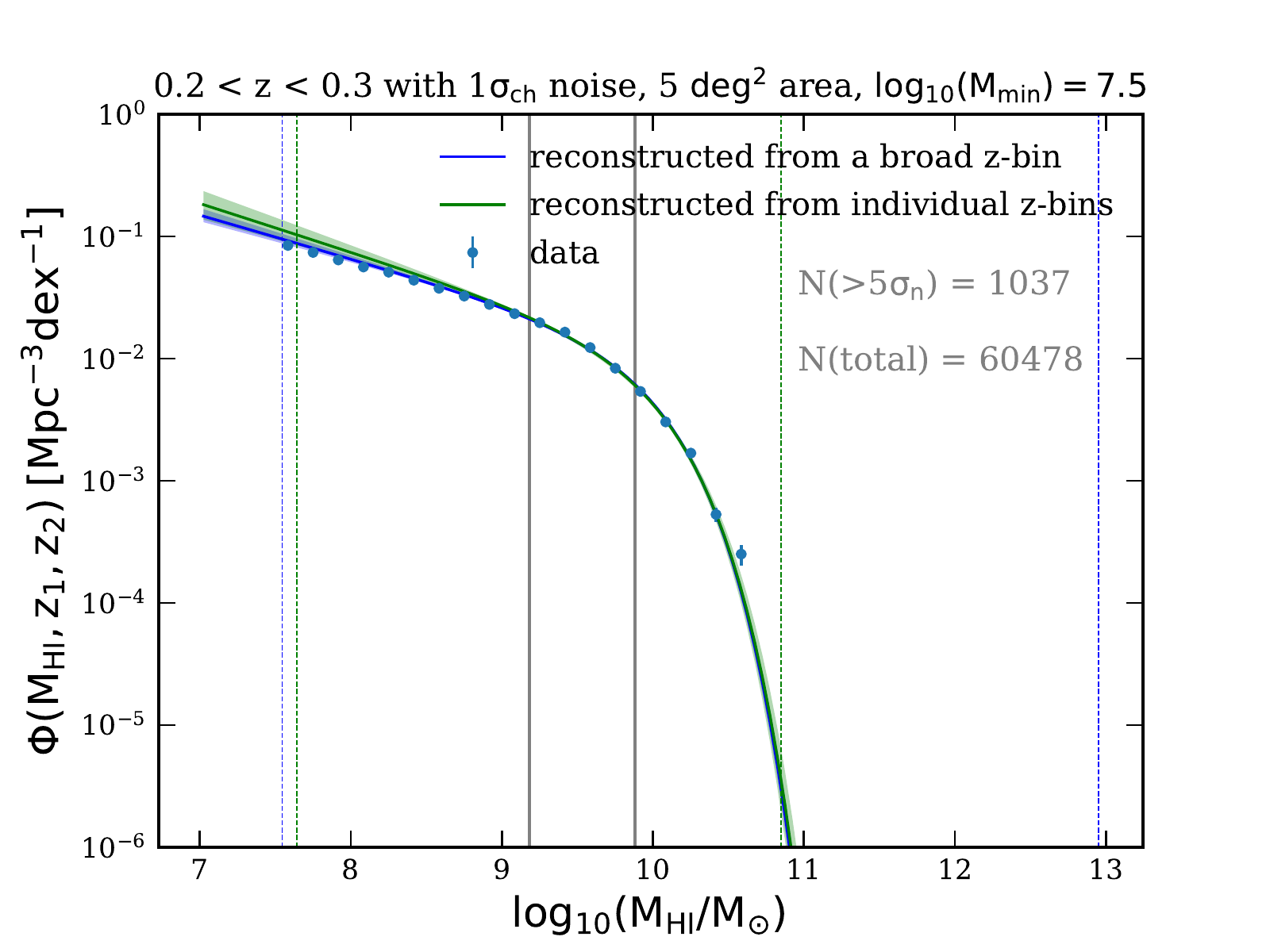}
  \end{subfigure}%
    \hfill
    \begin{subfigure}[b]{0.5\textwidth}
  \end{subfigure}%
  \hfill
  \raggedleft
  \begin{subfigure}[b]{0.5\textwidth}
    \includegraphics[width=1.06\columnwidth]{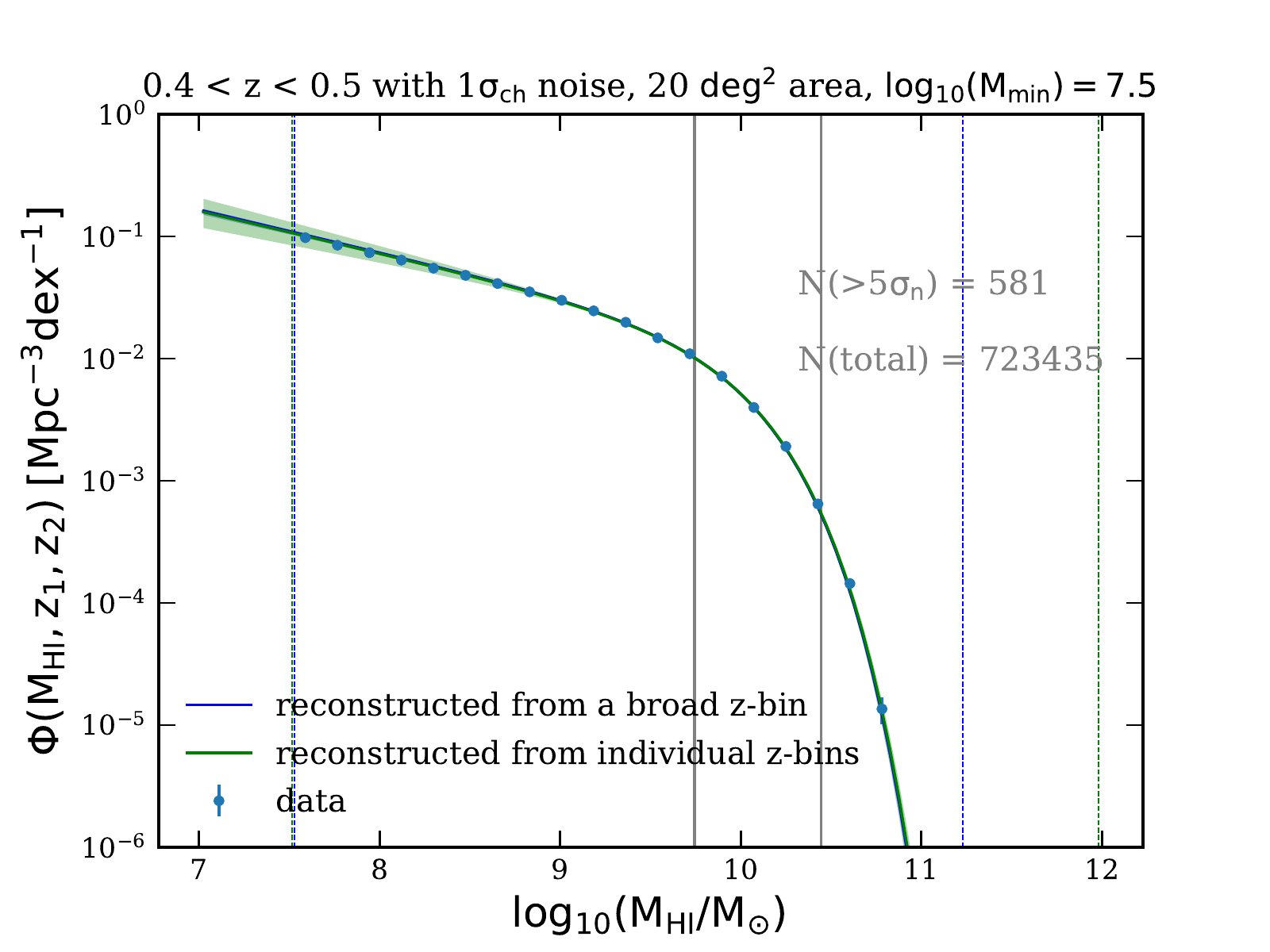}
  \end{subfigure}%
  \caption{Clockwise from top left to bottom right panels, the reconstructed HIMF from a single broad redshift bin (blue) and a smaller  redshift bin (green) from survey areas equal to 0.3, 5, 20\,deg${^2}$ in different redshift bins. The grey vertical lines indicate the $\sigma_{\rm n} = \sqrt{N_{\rm ch}} \sigma_{\rm ch} d\vi$ and 5\,$\sigma_{\rm n}$ detection threshold at the centre of each redshift bin. All other properties of this figure are the same as that of Fig.~\ref{fig:himf_s}. }
  \label{fig:himf_a}
\end{figure*}

Finally, we look at the effect of changing the survey area on the reconstructed model HIMF for our baseline survey. We show the posterior values for the reconstructed HIMF model parameters over a range of areas from 0.3 to 20 deg$^{2}$ in Table~\ref{tab:pos_a}. Fig.~\ref{fig:himf_a} shows the reconstructed HIMF over these different survey areas and its evolution with redshift.

\begin{figure*}
  \centering
  \begin{subfigure}[b]{0.5\textwidth}
    \includegraphics[width=1.06\columnwidth]{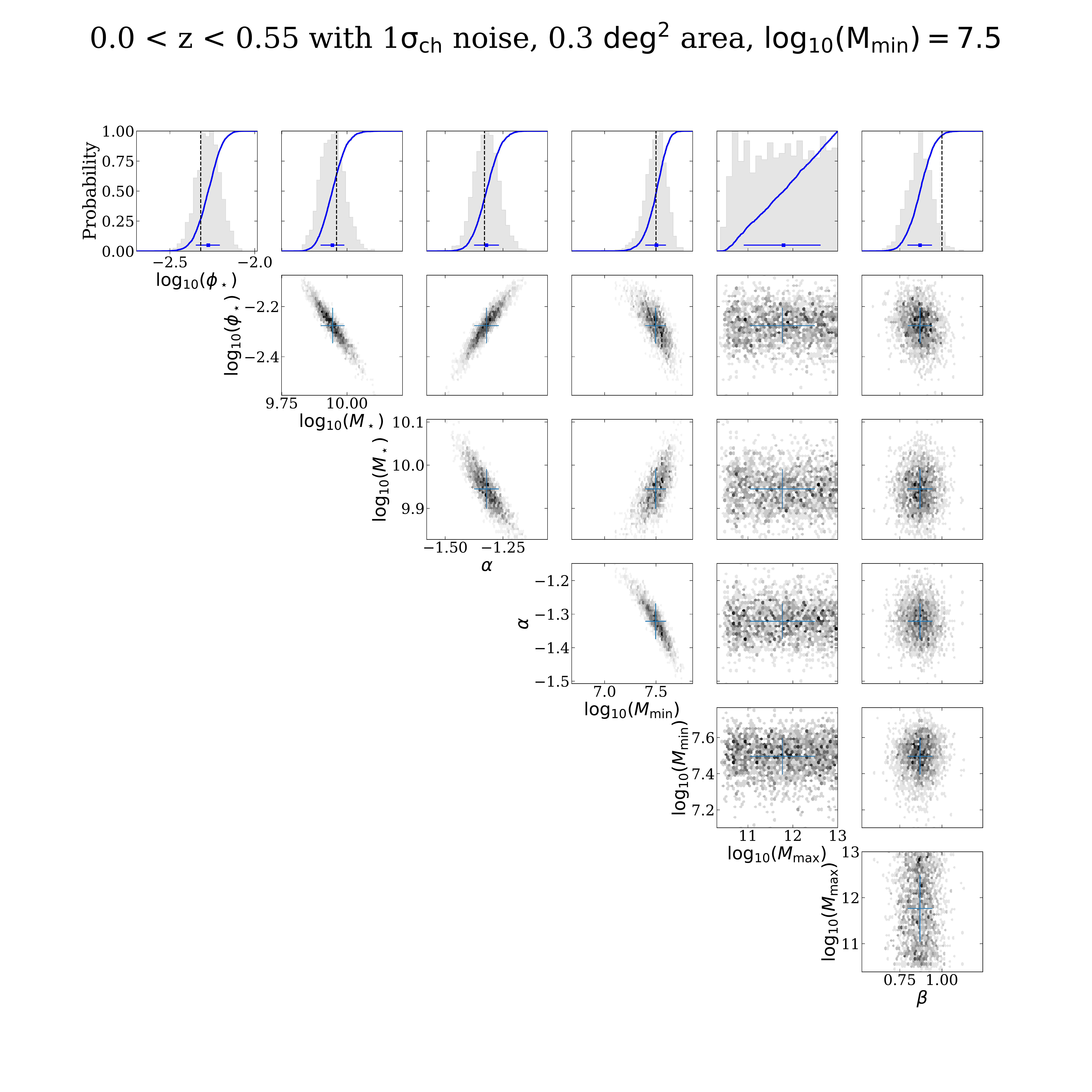}
  \end{subfigure}%
  \hfill
  \begin{subfigure}[b]{0.5\textwidth}
    \includegraphics[width=1.06\columnwidth]{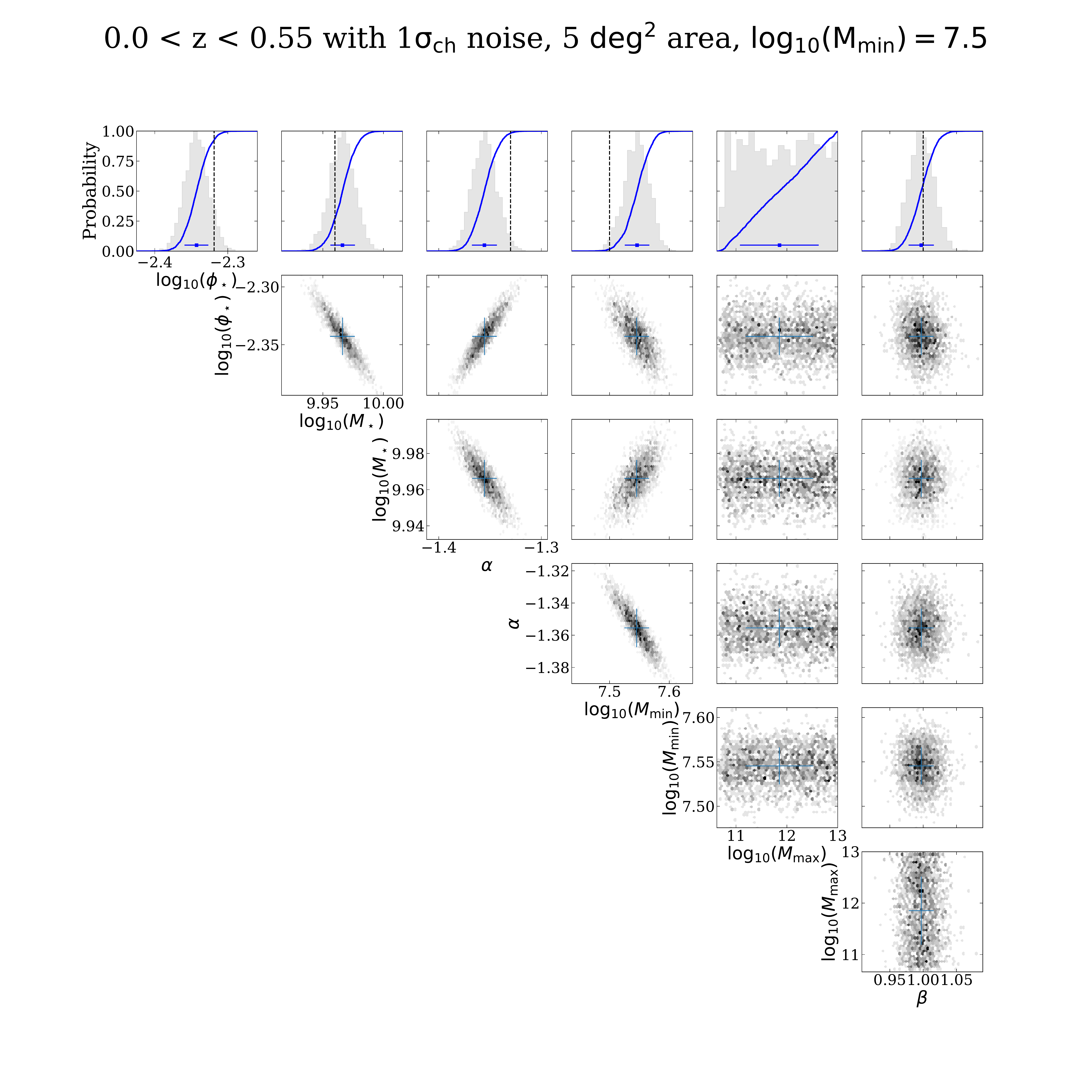}
  \end{subfigure}%
    \hfill
    \begin{subfigure}[b]{0.5\textwidth}
  \end{subfigure}%
  \hfill
  \raggedleft
  \begin{subfigure}[b]{0.5\textwidth}
    \includegraphics[width=1.06\columnwidth]{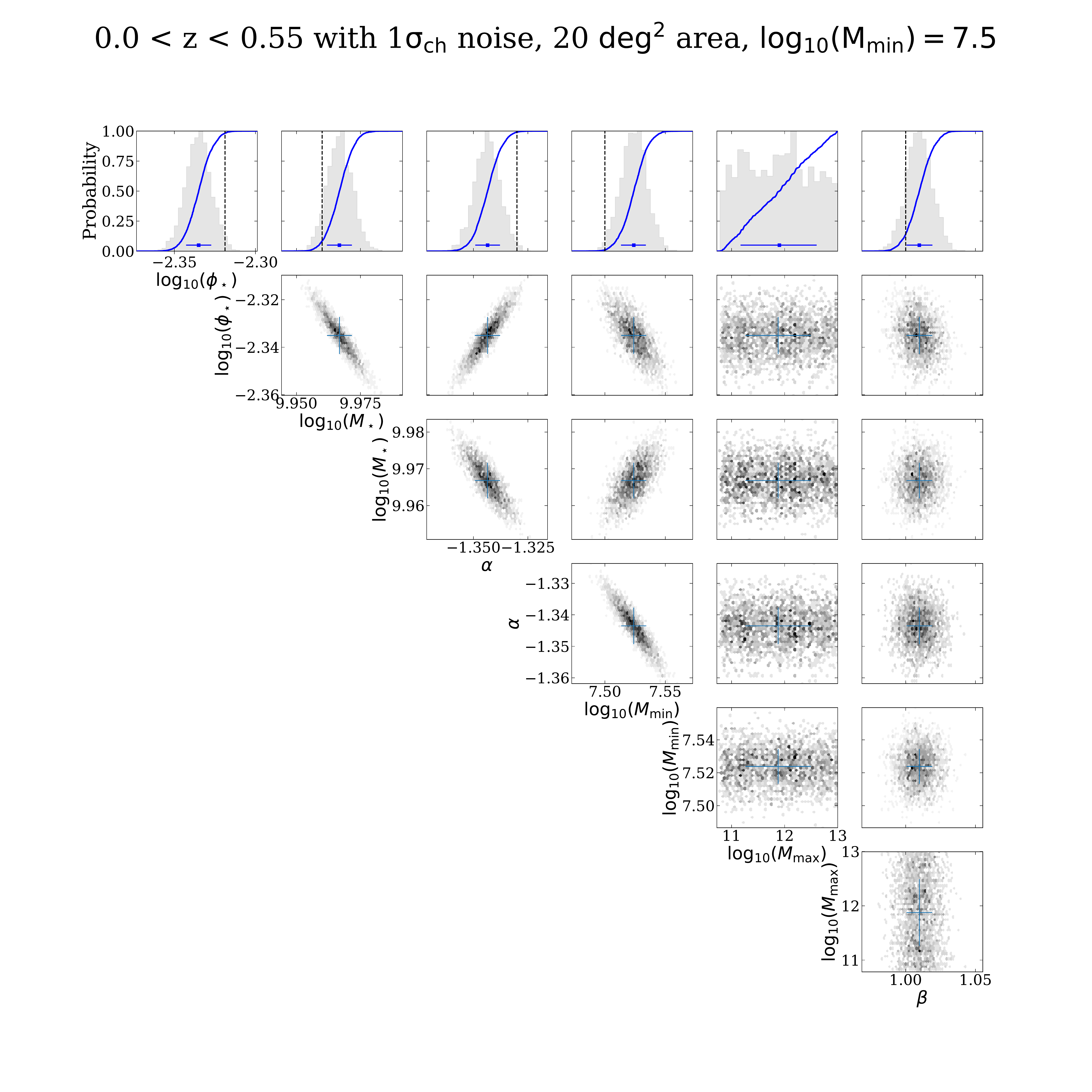}
  \end{subfigure}%
 \caption{The posterior distributions for the HIMF model parameters for a single broad redshift bin for survey areas equal to 0.3, 5, 20\,deg$^2$ clockwise from top left to bottom right panels. 
 All other properties of this figure are the same as that of Fig.~\ref{fig:marg_zbin}. }
 \label{fig:marg_a}
\end{figure*}

Fig.~\ref{fig:marg_a} shows the posterior distributions for the parameters of the HIMF model measured in a broad z-bin over 0.3 to 20 $ \mathrm{deg^2}$ areas. It indicates that all the parameters $\phi_\star, M_\star, \alpha, M_{\rm min}, M_{\rm max}, \beta$ and the \ha mass density $\Omega_{\rm HI}$  become better constrained as the survey area increases as expected. 

The key advantage in moving to larger area is the increase in volume at lower redshift, which enables one to fully constrain the mass function in smaller redshift bins, as there are enough galaxies at all masses. This is shown most clearly in Table~\ref{tab:pos_a} where the parameters that define the Schechter function are most closely aligned with their input values as area increases, even at $z<0.1$. 

\section{Conclusions}
\label{sec:conclusions}

We present a source-count model for the \ha flux distribution in galaxies and a Bayesian stacking technique to constrain the \ha mass function (HIMF), and its evolution over the redshift range $0 < z < 0.55$, below the detection threshold and down to $M_{\rm HI} = 10^{7.5}$ M$_{\odot}$ for this particular simulation.  We generate galaxy samples using an assumed HIMF model and simulate the \ha emission lines with different levels of background noise and different survey areas to investigate the robustness of this technique. We use the Monte-Carlo sampling algorithm {\sc Multinest} to reconstruct the HIMF model parameters, both for a single broad redshift bin and for individual redshift bins. In both cases we demonstrate that the HIMF can be accurately reconstructed from the simulated data using our method, which allows us to push the boundary of the HIMF far below the expected \ha mass detection limit of $M_{\rm HI} \sim 10^9 \mathrm{M_{\odot}}$ at $z\sim 0.1$ for the  MIGHTEE and LADUMA \ha surveys. More specifically, we find that:
\begin{itemize}

\item As expected, constraints on the Schechter function parameters $\phi_\star$, $M_\star$ and $\alpha$ become tighter as the background noise decreases, but the term that parameterises the evolution remains unchanged. Thus, if one has confidence that the shape of the HIMF does not change but only evolves as mass evolution, or in our case, pure density evolution, then this evolution can be measured robustly with relatively poor data. This is because once the low-mass slope has been well fit at low redshift, and the characteristic mass is well defined at intermediate redshift, then the evolution parameter cannot vary from its true value.  As such, the \ha mass density as a function of redshift can be measured robustly to high redshift, as this depends most strongly on the position of the knee in the mass function ($M_{\star}$).

\item The constraints on $\phi_\star$, $M_\star$ and $\alpha$, along with $\beta$ are weaker if the minimum mass is increased from $M_{\rm min} = 10^{7.5}$ to $10^{8.5}$\,M$_\odot$. However, our parameterisation means that we naturally find the minimum \ha mass ($M_{\rm min})$ that the data allows the model to fit to.

\item The constraints on the parameters $\phi_\star$, $M_\star$, $\alpha$ and $\beta$ become stronger as the survey area increases. However, increasing the area naturally leads to a more robust determination of the HIMF at the lowest redshifts due to the significant increase in volume that will hence contain the higher mass galaxies.

\end{itemize}

This work is a first step towards the measurement of the HIMF in the high redshift regime. In future work we will apply this method to real data from the MIGHTEE survey. However, there are some caveats one needs to be aware of in doing so: 
\begin{itemize}
\item In practice, the noise may not necessarily follow a Gaussian distribution, and thus we would need to determine the real noise distribution to be used in place of the Gaussian distribution in equation~\eqref{eq:salient}. This may lead to a significant increase in the computing time for equation~\eqref{eq:salient}, and especially if the $\sigma_{\rm n}$ is dependent on redshift.

\item We have also not considered the effect of confusion on our results, since we expect future deep \ha surveys to achieve $5-15$\,arcsec spatial resolution and a spectral resolution of at least a few 10\,km\,s$^{-1}$. However confusion will need to be considered if our approach is used on lower resolution data (e.g. HIPASS and ALFALFA).

\item We also assume that the galaxy sample is complete for a given mass threshold, in this case $M_{\rm min} = 10^{7.5}$ or $10^{8.5}$ M$_\odot$. However, in practice a complete mass-limited sample of galaxies is difficult to construct, especially at the higher redshifts. Thus, as with any other experiment that involves stacking based on a sample defined at a different wavelength, one always needs to be aware that the derived HIMF is that of the galaxies in the parent sample, and will always miss galaxies that could have relatively large \ha masses but low stellar masses.

\item Other potential issues are instrumental effects that generate spectral features below the $5\sigma$ noise level and might contaminate the flux in our Bayesian stacking signal, e.g. strong continuum sources and their sidelobes that are not subtracted accurately enough. Such contamination should not correlate with the galaxy catalogue used for stacking and so the stacked flux should be unbiased on average. However, the residuals will still contribute to the noise level and ultimately we may need to test different cleaning techniques or use independent noise estimates for each galaxy.

\item If we were interested in extracting the fluxes of sources which we expect to be extended based on their optical properties, we could enlarge the aperture over which we extract the flux. However, we note that one could also simply image the cube with a lower resolution restoring beam at the high-frequency (low-redshift) end of the data cube, ensuring that all sources are unresolved.

\end{itemize}

\section*{Acknowledgements}
We thank the anonymous referee for helpful comments that have contributed to this paper. This work was supported by the China Scholarship Council and the Science and Technology Facilities Council of the United Kingdom. MJJ and IH are supported by the South African Radio Astronomy Observatory and STFC under the consolidated grant ST/S000488/1. JRA acknowledges support from a Christ Church Career Development Fellowship. MGS acknowledges support from the South African Square Kilometre Array Project and National Research Foundation (Grant No. 84156). IH acknowledges support from the Oxford Hintze Centre for Astrophysical Surveys which is funded through generous support from the Hintze Family Charitable Foundation.

%%%%%%%%%%%%%%%%%%%%%%%%%%%%%%%%%%%%%%%%%%%%%%%%%%

\begin{table*}
	\centering
	\caption{Reconstructed parameters of the HIMF model from our baseline survey with increasing minimum mass of the \ha galaxy sample, where we fix $\log_{10}(\phi_\star) = -2.318$ for individual redshift bins. We note that $\Omega_{\rm HI}$ is estimated by integrating the HIMF down to $M_{\rm HI} = 10^{7.5}$\,M$_{\odot}$ for both cases.}
	\label{tab:pos_m}
	\begin{tabular}{lcrrrrrrr} 
	\hline
	\hline
	Parameter & $\log_{10}(M_{\rm min})$ & $0 < z < 0.1$ & $0.1 < z < 0.2$ & $0.2 < z < 0.3$ & $0.3 < z < 0.4$ & $0.4 < z < 0.5$ & $0.5 < z < 0.55$  & $0 < z < 0.55$\\
	 & (input) & &  &  &  &  &  & \\
	\hline
        $\log_{10}(\phi_\star)$ & 7.5 & -2.318 & -2.318 & -2.318 & -2.318 & -2.318 & -2.318 & -2.397$^{+0.039}_{-0.042}$ \\
$\log_{10}(\phi_\star)$ & 8.5 & -2.318 & -2.318 & -2.318 & -2.318 & -2.318 & -2.318 & -2.37$^{+0.055}_{-0.054}$ \\

	\hline
        $\log_{10}(M_\star)$ & 7.5 & 9.819$^{+0.075}_{-0.075}$ & 9.955$^{+0.051}_{-0.061}$ & 10.02$^{+0.039}_{-0.048}$ & 9.962$^{+0.071}_{-0.068}$ & 9.928$^{+0.088}_{-0.067}$ & 9.918$^{+0.089}_{-0.098}$ & 9.996$^{+0.027}_{-0.025}$ \\
$\log_{10}(M_\star)$ & 8.5 & 9.851$^{+0.089}_{-0.085}$ & 9.932$^{+0.062}_{-0.068}$ & 10.029$^{+0.029}_{-0.042}$ & 9.941$^{+0.08}_{-0.064}$ & 9.99$^{+0.088}_{-0.086}$ & 9.957$^{+0.107}_{-0.142}$ & 9.987$^{+0.032}_{-0.034}$ \\

	\hline
        $\alpha$ & 7.5 & -1.389$^{+0.041}_{-0.041}$ & -1.358$^{+0.053}_{-0.042}$ & -1.425$^{+0.056}_{-0.041}$ & -1.347$^{+0.108}_{-0.098}$ & -1.233$^{+0.166}_{-0.159}$ & -1.293$^{+0.245}_{-0.181}$ & -1.393$^{+0.028}_{-0.028}$ \\
$\alpha$ & 8.5 & -1.353$^{+0.087}_{-0.079}$ & -1.304$^{+0.09}_{-0.076}$ & -1.489$^{+0.059}_{-0.041}$ & -1.319$^{+0.127}_{-0.14}$ & -1.347$^{+0.177}_{-0.164}$ & -1.479$^{+0.423}_{-0.248}$ & -1.386$^{+0.051}_{-0.047}$ \\

	\hline
        $\log_{10}(M_{\rm min})$ & 7.5 & 7.56$^{+0.068}_{-0.066}$ & 7.544$^{+0.069}_{-0.082}$ & 7.636$^{+0.078}_{-0.096}$ & 7.487$^{+0.18}_{-0.255}$ & 7.169$^{+0.383}_{-0.943}$ & 7.337$^{+0.406}_{-1.189}$ & 7.58$^{+0.042}_{-0.049}$ \\
$\log_{10}(M_{\rm min})$ & 8.5 & 8.446$^{+0.046}_{-0.056}$ & 8.471$^{+0.035}_{-0.039}$ & 8.567$^{+0.032}_{-0.034}$ & 8.469$^{+0.076}_{-0.094}$ & 8.454$^{+0.108}_{-0.146}$ & 8.643$^{+0.121}_{-0.276}$ & 8.502$^{+0.023}_{-0.025}$ \\

	\hline
        $\log_{10}(M_{\rm max})$ & 7.5 & 11.46$^{+1.066}_{-1.129}$ & 11.666$^{+0.93}_{-0.932}$ & 11.773$^{+0.829}_{-0.8}$ & 11.709$^{+0.831}_{-0.861}$ & 11.712$^{+0.839}_{-0.84}$ & 11.783$^{+0.757}_{-0.852}$ & 11.794$^{+0.829}_{-0.819}$ \\
$\log_{10}(M_{\rm max})$ & 8.5 & 11.455$^{+1.056}_{-1.072}$ & 11.702$^{+0.904}_{-0.967}$ & 11.822$^{+0.799}_{-0.83}$ & 11.654$^{+0.886}_{-0.817}$ & 11.693$^{+0.883}_{-0.845}$ & 11.842$^{+0.799}_{-0.88}$ & 11.796$^{+0.845}_{-0.854}$ \\

	\hline
        $\beta$ & 7.5 & 1.142$^{+1.816}_{-1.493}$ & 0.591$^{+1.258}_{-1.019}$ & -0.173$^{+0.733}_{-0.579}$ & 0.75$^{+0.859}_{-0.892}$ & 1.346$^{+0.73}_{-0.951}$ & 1.098$^{+0.999}_{-0.938}$ & 0.968$^{+0.042}_{-0.039}$ \\
$\beta$ & 8.5 & 1.289$^{+1.734}_{-1.536}$ & 1.255$^{+1.529}_{-1.455}$ & -0.446$^{+0.71}_{-0.381}$ & 1.091$^{+0.794}_{-0.998}$ & 0.704$^{+0.86}_{-0.962}$ & 0.793$^{+1.429}_{-1.208}$ & 0.928$^{+0.074}_{-0.067}$ \\

	\hline
	\hline
	$\Omega_{\rm HI} \times 10^4 \;(\rm individual)$ & 7.5 & 3.7$^{+1.08}_{-0.79}$ & 5.01$^{+0.55}_{-0.39}$ & 5.64$^{+0.44}_{-0.34}$ & 5.85$^{+0.52}_{-0.41}$ & 6.25$^{+0.66}_{-0.53}$ & 6.57$^{+1.11}_{-1.04}$ &  \\
$\Omega_{\rm HI} \times 10^4 \;(\rm individual)$ & 8.5 & 3.87$^{+1.26}_{-0.95}$ & 5.0$^{+0.45}_{-0.48}$ & 6.1$^{+0.52}_{-0.5}$ & 6.0$^{+0.86}_{-0.57}$ & 6.87$^{+1.07}_{-1.12}$ & 8.68$^{+4.72}_{-2.66}$ &  \\

	\hline
	$\Omega_{\rm HI} \times 10^4 \;(\rm broad)$ & 7.5 & 4.74$^{+0.2}_{-0.18}$ & 5.11$^{+0.21}_{-0.19}$ & 5.52$^{+0.22}_{-0.19}$ & 5.94$^{+0.23}_{-0.19}$ & 6.36$^{+0.24}_{-0.2}$ & 6.67$^{+0.25}_{-0.22}$ &  \\
$\Omega_{\rm HI} \times 10^4 \;(\rm broad)$ & 8.5 & 4.9$^{+0.3}_{-0.31}$ & 5.26$^{+0.3}_{-0.28}$ & 5.66$^{+0.29}_{-0.26}$ & 6.07$^{+0.3}_{-0.27}$ & 6.49$^{+0.31}_{-0.3}$ & 6.79$^{+0.35}_{-0.32}$ &  \\

	\hline
	\end{tabular}
\end{table*}

\begin{table*}
	\centering
	\caption{Reconstructed parameters of the HIMF model from our baseline survey with various survey areas, where we fix $\log_{10}(\phi_\star) = -2.318$ for individual redshift bins.}
	\label{tab:pos_a}
	\begin{tabular}{lrrrrrrrr} 
	\hline
	\hline
	Parameter & Area [deg$^2$] & $0 < z < 0.1$ & $0.1 < z < 0.2$ & $0.2 < z < 0.3$ & $0.3 < z < 0.4$ & $0.4 < z < 0.5$ & $0.5 < z < 0.55$ & $0 < z < 0.55$\\
	\hline
        $\log_{10}(\phi_\star)$ & 0.3 & -2.318 & -2.318 & -2.318 & -2.318 & -2.318 & -2.318 & -2.274$^{+0.068}_{-0.073}$ \\
$\log_{10}(\phi_\star)$ & 1.0 & -2.318 & -2.318 & -2.318 & -2.318 & -2.318 & -2.318 & -2.397$^{+0.039}_{-0.042}$ \\
$\log_{10}(\phi_\star)$ & 5.0 & -2.318 & -2.318 & -2.318 & -2.318 & -2.318 & -2.318 & -2.343$^{+0.016}_{-0.017}$ \\
$\log_{10}(\phi_\star)$ & 20.0 & -2.318 & -2.318 & -2.318 & -2.318 & -2.318 & -2.318 & -2.335$^{+0.008}_{-0.008}$ \\

	\hline
        $\log_{10}(M_\star)$ & 0.3 & 10.08$^{+0.562}_{-0.162}$ & 9.994$^{+0.083}_{-0.08}$ & 9.908$^{+0.1}_{-0.093}$ & 9.787$^{+0.068}_{-0.048}$ & 9.947$^{+0.11}_{-0.082}$ & 9.917$^{+0.135}_{-0.123}$ & 9.944$^{+0.046}_{-0.045}$ \\
$\log_{10}(M_\star)$ & 1.0 & 9.819$^{+0.075}_{-0.075}$ & 9.955$^{+0.051}_{-0.061}$ & 10.02$^{+0.039}_{-0.048}$ & 9.962$^{+0.071}_{-0.068}$ & 9.928$^{+0.088}_{-0.067}$ & 9.918$^{+0.089}_{-0.098}$ & 9.996$^{+0.027}_{-0.025}$ \\
$\log_{10}(M_\star)$ & 5.0 & 9.973$^{+0.037}_{-0.038}$ & 9.97$^{+0.033}_{-0.029}$ & 9.986$^{+0.024}_{-0.025}$ & 9.962$^{+0.025}_{-0.026}$ & 9.967$^{+0.03}_{-0.031}$ & 9.98$^{+0.056}_{-0.055}$ & 9.966$^{+0.01}_{-0.01}$ \\
$\log_{10}(M_\star)$ & 20.0 & 9.972$^{+0.021}_{-0.025}$ & 9.96$^{+0.015}_{-0.014}$ & 9.967$^{+0.011}_{-0.011}$ & 9.978$^{+0.012}_{-0.012}$ & 9.972$^{+0.014}_{-0.014}$ & 9.963$^{+0.026}_{-0.024}$ & 9.967$^{+0.005}_{-0.005}$ \\

	\hline
        $\alpha$ & 0.3 & -1.278$^{+0.085}_{-0.064}$ & -1.356$^{+0.072}_{-0.061}$ & -1.272$^{+0.136}_{-0.119}$ & -1.012$^{+0.06}_{-0.116}$ & -1.181$^{+0.142}_{-0.201}$ & -1.148$^{+0.117}_{-0.139}$ & -1.321$^{+0.054}_{-0.053}$ \\
$\alpha$ & 1.0 & -1.389$^{+0.041}_{-0.041}$ & -1.358$^{+0.053}_{-0.042}$ & -1.425$^{+0.056}_{-0.041}$ & -1.347$^{+0.108}_{-0.098}$ & -1.233$^{+0.166}_{-0.159}$ & -1.293$^{+0.245}_{-0.181}$ & -1.393$^{+0.028}_{-0.028}$ \\
$\alpha$ & 5.0 & -1.342$^{+0.026}_{-0.021}$ & -1.345$^{+0.025}_{-0.025}$ & -1.398$^{+0.031}_{-0.029}$ & -1.376$^{+0.039}_{-0.039}$ & -1.332$^{+0.059}_{-0.059}$ & -1.375$^{+0.119}_{-0.113}$ & -1.356$^{+0.012}_{-0.012}$ \\
$\alpha$ & 20.0 & -1.345$^{+0.017}_{-0.012}$ & -1.34$^{+0.013}_{-0.013}$ & -1.342$^{+0.014}_{-0.014}$ & -1.375$^{+0.019}_{-0.019}$ & -1.342$^{+0.028}_{-0.025}$ & -1.307$^{+0.06}_{-0.053}$ & -1.343$^{+0.006}_{-0.006}$ \\

	\hline
        $\log_{10}(M_{\rm min})$ & 0.3 & 7.445$^{+0.13}_{-0.169}$ & 7.631$^{+0.102}_{-0.132}$ & 7.328$^{+0.241}_{-0.453}$ & 6.12$^{+0.802}_{-0.738}$ & 6.895$^{+0.626}_{-1.138}$ & 6.341$^{+0.732}_{-0.794}$ & 7.504$^{+0.092}_{-0.109}$ \\
$\log_{10}(M_{\rm min})$ & 1.0 & 7.56$^{+0.068}_{-0.066}$ & 7.544$^{+0.069}_{-0.082}$ & 7.636$^{+0.078}_{-0.096}$ & 7.487$^{+0.18}_{-0.255}$ & 7.169$^{+0.383}_{-0.943}$ & 7.337$^{+0.406}_{-1.189}$ & 7.58$^{+0.042}_{-0.049}$ \\
$\log_{10}(M_{\rm min})$ & 5.0 & 7.522$^{+0.031}_{-0.03}$ & 7.513$^{+0.036}_{-0.038}$ & 7.639$^{+0.045}_{-0.051}$ & 7.595$^{+0.069}_{-0.077}$ & 7.487$^{+0.124}_{-0.149}$ & 7.552$^{+0.2}_{-0.296}$ & 7.546$^{+0.021}_{-0.021}$ \\
$\log_{10}(M_{\rm min})$ & 20.0 & 7.502$^{+0.017}_{-0.016}$ & 7.529$^{+0.018}_{-0.02}$ & 7.531$^{+0.026}_{-0.027}$ & 7.584$^{+0.037}_{-0.039}$ & 7.511$^{+0.055}_{-0.067}$ & 7.377$^{+0.132}_{-0.177}$ & 7.524$^{+0.01}_{-0.011}$ \\

	\hline
        $\log_{10}(M_{\rm max})$ & 0.3 & 10.75$^{+1.49}_{-0.97}$ & 11.658$^{+0.898}_{-1.058}$ & 11.676$^{+0.873}_{-0.919}$ & 11.654$^{+0.879}_{-0.878}$ & 11.814$^{+0.775}_{-0.865}$ & 11.525$^{+0.978}_{-0.786}$ & 11.793$^{+0.824}_{-0.888}$ \\
$\log_{10}(M_{\rm max})$ & 1.0 & 11.46$^{+1.066}_{-1.129}$ & 11.666$^{+0.93}_{-0.932}$ & 11.773$^{+0.829}_{-0.8}$ & 11.709$^{+0.831}_{-0.861}$ & 11.712$^{+0.839}_{-0.84}$ & 11.783$^{+0.757}_{-0.852}$ & 11.794$^{+0.829}_{-0.819}$ \\
$\log_{10}(M_{\rm max})$ & 5.0 & 11.642$^{+0.912}_{-0.936}$ & 11.754$^{+0.811}_{-0.902}$ & 11.796$^{+0.845}_{-0.802}$ & 11.882$^{+0.757}_{-0.902}$ & 11.821$^{+0.796}_{-0.824}$ & 11.895$^{+0.726}_{-0.769}$ & 11.855$^{+0.77}_{-0.777}$ \\
$\log_{10}(M_{\rm max})$ & 20.0 & 11.633$^{+0.871}_{-0.939}$ & 11.739$^{+0.864}_{-0.859}$ & 11.835$^{+0.765}_{-0.739}$ & 11.808$^{+0.792}_{-0.721}$ & 11.924$^{+0.749}_{-0.777}$ & 11.932$^{+0.732}_{-0.776}$ & 11.9$^{+0.701}_{-0.732}$ \\

	\hline
        $\beta$ & 0.3 & 1.746$^{+1.552}_{-1.778}$ & 1.271$^{+1.663}_{-1.516}$ & 1.636$^{+1.361}_{-1.467}$ & 3.077$^{+0.521}_{-0.749}$ & 1.358$^{+0.791}_{-1.195}$ & 1.029$^{+1.054}_{-1.12}$ & 0.87$^{+0.071}_{-0.075}$ \\
$\beta$ & 1.0 & 1.142$^{+1.816}_{-1.493}$ & 0.591$^{+1.258}_{-1.019}$ & -0.173$^{+0.733}_{-0.579}$ & 0.75$^{+0.859}_{-0.892}$ & 1.346$^{+0.73}_{-0.951}$ & 1.098$^{+0.999}_{-0.938}$ & 0.968$^{+0.042}_{-0.039}$ \\
$\beta$ & 5.0 & 0.638$^{+1.545}_{-1.154}$ & 0.651$^{+0.605}_{-0.676}$ & 0.389$^{+0.386}_{-0.383}$ & 0.747$^{+0.329}_{-0.324}$ & 0.925$^{+0.347}_{-0.359}$ & 0.684$^{+0.582}_{-0.615}$ & 0.997$^{+0.019}_{-0.019}$ \\
$\beta$ & 20.0 & 0.026$^{+1.046}_{-0.713}$ & 0.908$^{+0.293}_{-0.32}$ & 0.899$^{+0.17}_{-0.172}$ & 0.663$^{+0.156}_{-0.164}$ & 0.877$^{+0.164}_{-0.15}$ & 0.94$^{+0.264}_{-0.272}$ & 1.01$^{+0.009}_{-0.009}$ \\

	\hline
	\hline
	$\Omega_{\rm HI} \times 10^4 \;(\rm individual)$ & 0.3 & 4.21$^{+4.14}_{-1.1}$ & 5.76$^{+1.31}_{-0.76}$ & 5.38$^{+0.75}_{-0.5}$ & 6.01$^{+0.66}_{-0.66}$ & 6.54$^{+0.74}_{-1.04}$ & 5.47$^{+1.45}_{-1.3}$ &  \\
$\Omega_{\rm HI} \times 10^4 \;(\rm individual)$ & 1.0 & 3.68$^{+1.08}_{-0.81}$ & 5.0$^{+0.54}_{-0.4}$ & 5.58$^{+0.41}_{-0.31}$ & 5.85$^{+0.43}_{-0.37}$ & 6.29$^{+0.6}_{-0.52}$ & 6.39$^{+0.98}_{-0.81}$ &  \\
$\Omega_{\rm HI} \times 10^4 \;(\rm individual)$ & 5.0 & 4.96$^{+0.65}_{-0.57}$ & 5.2$^{+0.21}_{-0.21}$ & 5.66$^{+0.14}_{-0.15}$ & 5.98$^{+0.18}_{-0.16}$ & 6.52$^{+0.22}_{-0.26}$ & 6.71$^{+0.35}_{-0.46}$ &  \\
$\Omega_{\rm HI} \times 10^4 \;(\rm individual)$ & 20.0 & 4.72$^{+0.35}_{-0.31}$ & 5.22$^{+0.1}_{-0.1}$ & 5.71$^{+0.08}_{-0.08}$ & 6.05$^{+0.09}_{-0.09}$ & 6.53$^{+0.12}_{-0.12}$ & 6.62$^{+0.19}_{-0.21}$ &  \\

	\hline
	$\Omega_{\rm HI} \times 10^4 \;(\rm broad)$ & 0.3 & 5.15$^{+0.3}_{-0.44}$ & 5.5$^{+0.3}_{-0.43}$ & 5.88$^{+0.31}_{-0.42}$ & 6.27$^{+0.32}_{-0.42}$ & 6.65$^{+0.37}_{-0.44}$ & 6.93$^{+0.4}_{-0.47}$ &  \\
$\Omega_{\rm HI} \times 10^4 \;(\rm broad)$ & 1.0 & 4.72$^{+0.2}_{-0.17}$ & 5.09$^{+0.19}_{-0.18}$ & 5.5$^{+0.2}_{-0.18}$ & 5.91$^{+0.2}_{-0.18}$ & 6.34$^{+0.21}_{-0.2}$ & 6.64$^{+0.22}_{-0.22}$ &  \\
$\Omega_{\rm HI} \times 10^4 \;(\rm broad)$ & 5.0 & 4.78$^{+0.11}_{-0.07}$ & 5.16$^{+0.11}_{-0.07}$ & 5.58$^{+0.11}_{-0.06}$ & 6.01$^{+0.11}_{-0.06}$ & 6.45$^{+0.12}_{-0.06}$ & 6.77$^{+0.13}_{-0.06}$ &  \\
$\Omega_{\rm HI} \times 10^4 \;(\rm broad)$ & 20.0 & 4.83$^{+0.04}_{-0.04}$ & 5.21$^{+0.04}_{-0.04}$ & 5.65$^{+0.04}_{-0.05}$ & 6.1$^{+0.04}_{-0.05}$ & 6.54$^{+0.05}_{-0.05}$ & 6.88$^{+0.05}_{-0.05}$ &  \\

	\hline
	\end{tabular}
\end{table*}



\bibliographystyle{mnras}
\bibliography{references} % if your bibtex file is called example.bib
%%%%%%%%%%%%%%%%%%%%%%%%%%%%%%%%%%%%%%%%%%%%%%%%%%

%%%%%%%%%%%%%%%%%%%%%%%%%%%%%%%%%%%%%%%%%%%%%%%%%%
%%%%%%%%%%%%%%%%% APPENDICES %%%%%%%%%%%%%%%%%%%%%

%\appendix

%%%%%%%%%%%%%%%%%%%%%%%%%%%%%%%%%%%%%%%%%%%%%%%%%%

% Don't change these lines
\bsp	% typesetting comment
\label{lastpage}
\end{document}